\documentclass[10pt,prfluids,aps,showpacs,onecolumn,preprintnumbers,amsmath,amssymb,superscriptaddress,longbibliography]{revtex4-2}

\usepackage{graphicx}    
\usepackage{latexsym}    %
\usepackage{color}       %
\usepackage{dcolumn}     
\usepackage{bm}          
\usepackage{amssymb}     
\usepackage{hyperref}    
\expandafter\let\csname equation*\endcsname\relax
\expandafter\let\csname endequation*\endcsname\relax
\usepackage{amsmath}
\usepackage{ulem}
\usepackage{mathtools}          

\usepackage{graphics}
\usepackage{epsfig}
\usepackage{subfigure}
\usepackage{nicefrac}
\usepackage{verbatim}
\usepackage{enumerate}
\usepackage{footnote}
\usepackage{booktabs}
\usepackage{siunitx}
\usepackage{xcolor}

\DeclareMathOperator\erf{erf}
\DeclareMathOperator\erfc{erfc}

\begin{document}

\title{Single inertial particle statistics in turbulent flows from Lagrangian velocity models}
\author{Jan~Friedrich}
\affiliation{Institute of Physics and For Wind, University of Oldenburg, 26129 Oldenburg, Germany}
\affiliation{Univ. Lyon, ENS de Lyon, Univ. Claude Bernard, CNRS, Laboratoire de Physique,
F-69342, Lyon, France}
\author{Bianca~Viggiano}
\affiliation{Univ. Lyon, ENS de Lyon, Univ. Claude Bernard, CNRS, Laboratoire de Physique,
F-69342, Lyon, France}
\affiliation{Department of Mechanical and Materials Engineering, Portland State University, Portland,
Oregon, USA}
\author{Mickael~Bourgoin}
\affiliation{Univ. Lyon, ENS de Lyon, Univ. Claude Bernard, CNRS, Laboratoire de Physique,
F-69342, Lyon, France}
\author{Ra\'ul~Bayo\'an~Cal}
\affiliation{Department of Mechanical and Materials Engineering, Portland State University, Portland,
Oregon, USA}
\affiliation{Univ. Lyon, ENS de Lyon, Univ. Claude Bernard, CNRS, Laboratoire de Physique,
F-69342, Lyon, France}
\author{Laurent~Chevillard}
\affiliation{Univ. Lyon, ENS de Lyon, Univ. Claude Bernard, CNRS, Laboratoire de Physique,
F-69342, Lyon, France}
\date{\today}
\begin{abstract}
We present the extension of a modeling technique for Lagrangian tracer particles [B. Viggiano et al., J. Fluid Mech.(2020), vol. 900, A27] which accounts for the effects of particle inertia. Thereby, the particle velocity for several Stokes numbers is modeled directly by a multi-layered Ornstein-Uhlenbeck process and a comparison of key statistical quantities (second-order velocity structure function, acceleration correlation function, and root mean square acceleration) to expressions derived from Batchelor's model as well as to direct numerical simulations (DNS) is performed. In both approaches, Stokes' drag is treated by an approximate ``linear filter'' which replaces the particle position entering the fluid velocity field
by the corresponding ideal tracer position. Effects of preferential concentration of inertial particles are taken into account in terms of an effective Stokes number that is determined from the zero crossing of the acceleration correlation function from DNS. This approximation thus allows the modeling of inertial particle statistics through stochastic methods and models for the Lagrangian velocity; the particle velocity is effectively decoupled from the particle position. In contrast to the ordinary filtering technique [Cencini et al., J. Turbul. (2006), 7, N36], our method captures the effects of preferential concentration of particles at low Stokes numbers which manifest themselves for instance by a sharp decrease of the acceleration variance for increasing Stokes numbers.
\end{abstract}

\maketitle                             

\section{Introduction}
\label{sec:intro}
Transport processes of particles in turbulent flows play an important role in turbulence theory~\cite{monin,frisch:1995,Pope} and are
intimately related to the problems of turbulent mixing and turbulent diffusion.
Recent advances in particle tracking~\cite{toschi2009lagrangian,Machicoane2019} allow for the accurate quantification of single-particle statistics and yield important insights into the behavior of marked fluid particles (so-called Lagrangian tracers) as well as  heavier particles, i.e., particles with finite inertia~\cite{qureshi2008acceleration,ayyalasomayajula2006lagrangian}.
For such inertial particles, for instance, numerical and experimental evidence suggests the segregation of particles into clusters~\cite{Squires1991,Eaton1994,Monchaux2012,Bourgoin2014}. This phenomenon, generally known as \emph{preferential concentration}, is sometimes explained by particles evading dominant vortical structures and accumulating in flow regions of high strain~\cite{Maxey1983} although other mechanisms have been proposed~\cite{Chen2006,Gustavsson2016}.  A comprehensive understanding of these mechanisms, however, is complicated by the random and multiscale structure of the turbulent velocity field. On the other hand, such particle-laden flows are commonly encountered in many industrial and environmental processes, which creates an abundance of research devoted to characterizing their dynamical properties~\cite{bec2006acceleration,bourgoin2006role}.
Despite numerous experimental, numerical, and theoretical works~\cite{toschi2009lagrangian,Bourgoin2014,Gustavsson2016}, simple generic models which are capable to quantitatively reproduce basic dynamical Lagrangian properties (e.g., velocity and acceleration statistics) as well as the combined role of particle inertia and preferential concentration are still lacking.

Turbulent flow properties can be described either in a fixed frame of reference, the so-called Eulerian description, or alternatively by the Lagrangian description, where evolution of the flow is observed temporally through the trajectories of point particles. The Lagrangian description in turbulence provides a complete view of particle transport and dispersion which can be traced back to the seminal works by G.I. Taylor who set the diffusion problem in the context of fluid element trajectories~\cite{taylor:1922}.
Thereby, the trajectory of an ideal tracer, the so-called Lagrangian path, can be determined from the first-order ordinary differential equation (ODE)
\begin{equation}
  \mathbf{\dot X}(\mathbf{y},t)\equiv \mathbf{v}(\mathbf{y},t)=\mathbf{u}(\mathbf{X}(\mathbf{y},t),t)\;.
  \label{eq:vel_lag}
\end{equation}
where $\mathbf{v}(\mathbf{y},t)$  is the Lagrangian velocity, with the initial condition $\mathbf{X}(\mathbf{y},0)=\mathbf{y}$.
The statistical description of ideal tracer particles in turbulent flows, which is governed by Eq. (\ref{eq:vel_lag}), is one of the main concerns of turbulence theory. On the basis of the self-similar theory of Kolmogorov, moments of the Lagrangian velocity increments $\delta_{\tau} \mathbf{v}(\mathbf{y},t)= \mathbf{v}(\mathbf{y},t+\tau)-\mathbf{v}(\mathbf{y},t)$ are supposed to scale as $\langle (\delta_{\tau}v)^n \rangle  \sim \tau^{n/2}$.
Hence, Kolmogorov's theory suggests that a tracer particle exhibits a diffusive process (ordinary Brownian motion) in velocity space~\cite{monin,OBUKHOV1959113} at inertial scales. The corresponding Ornstein-Uhlenbeck (OU) process for the Lagrangian velocity has been extended by Sawford in order to account for finite Reynolds number effects~\cite{sawford1991reynolds} (see also~\cite{Pinton} for further references). Recently, an infinitely differentiable causal random walk has been proposed~\cite{Viggiano_2020}. It can be considered as a layered OU processes, where each layer of order $n$ ensures the existence of the derivative of order $n$ and causality. Nonetheless, experimental and numerical evidence suggests that moments exhibit multifractal scaling
$\langle (\delta_{\tau}v)^n \rangle  \sim \tau^{\zeta_n}$ which manifests itself in form of a nonlinear dependence of the scaling exponents $\zeta_n$ on order $n$~\cite{chevillard2003lagrangian,biferale2004multifractal,arneodo2008universal,chevillard2012phenomenological}. The latter feature is a direct consequence of the phenomenon of intermittency, reminiscent of the non-self-similarity of turbulence, which entails strongly non-Gaussian behavior of the Lagrangian velocity increment probability density function at small time separations $\tau$. Precisely those multiscaling (non-Gaussian) features impose a major challenge for the modeling
of particle transport in turbulent flows~\cite{friedrich:2003,Viggiano_2020}.

Once a particle gains inertia, either from its size or its density in comparison to the surrounding fluid with kinematic viscosity $\nu$, the dynamics of the particle is governed by additional forces~\cite{Gatignol,Maxey1983}. For example, there are now effects due to the Stokes' drag force based on the relative velocities of the fluid and the particles and an added mass force which is caused by the displacement of the fluid from the motion of the particle, to name a few.
If we only retain Stokes' drag and disregard these other terms (which is typically assumed to be the case when particles are rather heavy and small), the particle's velocity is determined by the following set of equations~\cite{Gatignol,Maxey1983,Pumir2016}
\begin{align}\label{eq:x_p}
  \mathbf{\dot X}_p(\mathbf{y},t)=& \mathbf{v}_p(\mathbf{y},t)\;, \\
  \mathbf{\dot v}_p(\mathbf{y},t)=&\frac{1}{\tau_p} \left [\mathbf{u}(\mathbf{X}_p(\mathbf{y},t),t)-\mathbf{v}_p(\mathbf{y},t) \right ]\;,
  \label{eq:v_p}
\end{align}
where $\mathbf{X}_p(\mathbf{y},t)$ is the particle position, $\mathbf{v}_p(\mathbf{y},t)$ is the particle velocity, and $\mathbf{u}(\mathbf{x},t)$ is the fluid velocity field. Initial conditions fully determine the future evolution of the particle position and we assume them to be given by $\mathbf{X}_p(\mathbf{y},0)=\mathbf{y}$ and $\mathbf{v}_p(\mathbf{y},0)= \mathbf{u}(\mathbf{y},0)$, where the matching of tracer and
inertial particle position at $t=0$ has been imposed arbitrarily. The particle response time, $\tau_p$, is defined as $\tau_p = m_p/3\pi \mu_f d_p$ where $m_p$ and $d_p$ denote the mass and the diameter of the particle, respectively, and $\mu_f$ is the viscosity of the fluid.

In this simplified model, the Stokes number, defined as St$=\tau_p/\tau_K$ (where $\tau_K=\left(\nu/\langle \varepsilon \rangle \right)^{1/2}$ is the Kolmogorov time scale and $\langle \varepsilon \rangle$ denotes the averaged local energy dissipation rate) is the relevant parameter to characterize particles inertia. The Stokes number thus describes the ratio of the particle response time $\tau_p$ to the Kolmogorov time scale $\tau_K$ at which viscous forces dominate the flow. In particular, for St$=0$ (i.e., for the case of an overdamped inertial particle) we recover Eq. (\ref{eq:vel_lag}) and the particle behaves as an ideal Lagrangian tracer.

The main purpose of this paper is to generalize recent stochastic and multifractal modeling approaches~\cite{Viggiano_2020} - which were initially devised for the Lagrangian velocity (\ref{eq:vel_lag}) - in order to include finite inertia effects. The simplest approach in this direction is to investigate the particle dynamics under the assumption that each inertial particle
samples homogeneously the carrier flow, and hence the explored Lagrangian flow velocity $\mathbf{u}(\mathbf{X}_p(\mathbf{y},t),t)$ in Eq. (\ref{eq:v_p}) is representative of the
complete background turbulence regardless of any eventual preferential concentration mechanism.
This approximation, determines the particle velocity $\mathbf{v}_p(\mathbf{y},t)$ by a \emph{linear filter} of the Lagrangian velocity $\mathbf{v}(\mathbf{y},t)$ (see Eq. (\ref{eq:v_p}))
and has been invoked for the first time by Tchen~\cite{chan2013mean}
and Hinze~\cite{hinze1959turbulence}. In particular, under the assumption of an exponential decay of the Lagrangian velocity correlation function for the tracers of the carrier flow (what would correspond for instance to a simple OU dynamics for the tracers), the Tchen-Hinze theory relates the variance of the particle velocity $\sigma_{\textbf{v}_p}^2$ directly to that of the ideal tracer $\sigma_{\textbf{v}}^2$ according to
\begin{equation}
    \sigma_{\textbf{v}_p}^2= \frac{T_L^2}{T_L^2 - \tau_p^2} \sigma_{\textbf{v}}^2 \;,
    \label{eq:tchen}
\end{equation}
where $T_L$ denotes the Lagrangian integral time scale. An extension of this linear filter approximation has been suggested by Deutsch and Simonin~\cite{deutsch1992dispersion,deutsch1991large} who argued that fluid quantities in Eq. (\ref{eq:tchen}) do not take into account the fact that the Lagrangian dynamics of the tracers probed at the position of the inertial particles may deviate from the global Lagrangian dynamics of the carrier flow, due to preferential concentration. Hence, they propose to replace the Lagrangian integral time scale $T_L$ and variance $\sigma_{\textbf{v}}^2$ by quantities determined \emph{at
the position of inertial particles} ($T_{L@p}$ and $\sigma_{\textbf{v}@p}^2$).
This refinement improves the agreement of Eq. (\ref{eq:tchen}) with direct numerical simulations~\cite{fevrier2000etude}, in particular for particles with small Stokes numbers, which seem to be more sensitive to preferential sampling effects. From a practical point of view, such a refinement requires however to determine $T_{L@p}$ and $\sigma_{\textbf{v}@p}^2$ on the basis of statistics of trajectories of individual inertial particle and its coincident Lagrangian particle, which is a quite challenging task, hardly feasible in experiments. Furthermore, the framework by Tchen-Hinze as well as the refined theory~\cite{deutsch1992dispersion,deutsch1991large} does not reproduce the observed behavior of statistical quantities at small time lags. This applies in particular to the particle's acceleration which remains non-differentiable under the assumption of a simple exponential correlation function for the Lagrangian velocity, in contrast to empirical evidence which shows smooth behavior. These limitations have also been discussed recently in the broader context of particle settling in turbulent flows~\cite{berk2021dynamics}.

In this context, the purpose of the present work thus is twofold: \emph{i.)} we generalize the Tchen-Hinze theory to a stochastic process which is infinitely differentiable (similarly to the framework which has been devised for the Lagrangian velocity~\cite{Viggiano_2020}) and, hence, contains important information on small-scale fluctuations, and, \emph{ii.)} we take into account the effects of preferential concentration (e.g., the correct amplitudes of particle accelerations~\cite{cencini2006dynamics}) by the introduction of an effective Stokes number, which is determined on the basis of the zero crossing of the empirically determined acceleration correlation function.

The paper is organized as follows: Section~\ref{sec:lin_filt} discusses the implications of the linear filter approximation for the particle velocity correlation function. Moreover, we provide a comparison between trajectories obtained under this approximation, and the ones obtained from direct numerical simulations (DNS) of turbulence. We present then models for particles correlation functions based on filtered versions of  \emph{i.)} an infinitely differentiable causal random walk in Section~\ref{sec:stoch_model} and \emph{ii.)} by the Batchelor model in Section~\ref{sec:batchelor}. These predictions are directly compared to DNS in Section~\ref{sec:DNScompare} and a summary of our results is included in Section~\ref{sec:conclusion}.

\section{Particle response based on the linear filter approximation}
\label{sec:lin_filt}
Inertial particle motion at low Stokes numbers, as mentioned in the introduction, is determined by the system of first order ODEs (\ref{eq:x_p}-\ref{eq:v_p}) and requires the knowledge of the full spatiotemporal (Eulerian)
fluid velocity field in Eq. (\ref{eq:v_p}). In the following, an approximation will be discussed, which replaces the Eulerian fluid velocity at the position of the inertial particle $\mathbf{X}_p(\mathbf{y},t)$ by its value at the position of the tracer, i.e. by the Lagrangian velocity. In this approximation, particle velocity is obtained by the linear filtering of tracer velocity.
\subsection{Linear filtering of Lagrangian velocity}
In order to model inertial particle statistics on the basis of the Lagrangian velocity (\ref{eq:vel_lag}), an approximation of the coupled system of first-order ODEs is invoked which can be termed ``linear filtering of the particle velocity'' due to its analogy to methods from signal processing~\cite{cencini2006dynamics}. In this approximation, the particle position $\mathbf{X}_p(\mathbf{y},t)$ that enters the fluid velocity field in Eq. (\ref{eq:v_p}) is approximated as the position of the ideal tracer $\mathbf{X}(\mathbf{y},t)$ whose temporal evolution is governed by equation (\ref{eq:vel_lag}). Therefore, Eqs.
(\ref{eq:x_p}-\ref{eq:v_p}) are approximated by
\begin{align}\label{eq:x_p_filter}
  \mathbf{\dot X}_p(\mathbf{y},t)=& \mathbf{v}_p(\mathbf{y},t)\;, \\
  \mathbf{\dot v}_p(\mathbf{y},t)=&\frac{1}{\tau_p} \left [\mathbf{u}(\mathbf{X}(\mathbf{y},t),t)-\mathbf{v}_p(\mathbf{y},t) \right ]\;.
  \label{eq:v_p_filter}
\end{align}
Hence, in this linear filter approximation, the particle velocity is effectively decoupled from the particle position. Therefore, the temporal evolution of the particle velocity is solely determined by the Lagrangian velocity $\mathbf{v}(\mathbf{y},t)$ along the tracer trajectory starting from the initial position of the inertial particle $\mathbf{X}_p(\mathbf{y},0)=\mathbf{y}$. To some extent, the linear filter neglects the spatiotemporal organization of the fluid velocity, and thus the segregation of
inertial particles in regions of low vorticity, as suggested by the phenomenon of preferential concentration.
Nevertheless, under this approximation, the evolution equation for the particle velocity (\ref{eq:v_p_filter}) can be solved according to
\begin{equation}
  \mathbf{v}_p(\mathbf{y},t) = \mathbf{v}(\mathbf{y},0) e^{-t/\tau_p}+ \frac{1}{\tau_p}\int_0^t \textrm{d}t' e^{-(t-t')/\tau_p} \mathbf{v}(\mathbf{y},t')\;.
\end{equation}
The particle position is thus determined as
\begin{equation}
  \mathbf{X}_p(\mathbf{y},t) = \mathbf{y}+ \tau_p \mathbf{v}(\mathbf{y},0)
  (1- e^{-t/\tau_p})+ \frac{1}{\tau_p}\int_0^t \textrm{d}t' \int_0^{t'} \textrm{d}t'' e^{-(t'-t'')/\tau_p} \mathbf{v}(\mathbf{y},t'')\;.
\end{equation}
The implications of this linear filter approximation for particle velocity and acceleration correlation functions will be discussed in the following section.

\subsection{Linear filtering of correlation functions}
In this section, the linear filter approximation (\ref{eq:v_p_filter}) is applied in order to establish relations between inertial particle velocity correlation functions and the Lagrangian velocity correlation function.
To this end, without loss of generality, we denote $v_p$ as of the component of the particle velocity vector, and correspondingly $a_p$ any component of the particle acceleration vector. Furthermore, we assume that the particle velocity has reached a statistically stationary state, such that we can neglect the dependence on its initial position. Eq. (\ref{eq:v_p_filter}) can then be integrated according to
\begin{align}
v_p(t)= \frac{1}{\tau_p}\int_{-\infty}^t \text{d}t' e^{-(t-t')/\tau_p}u(\boldsymbol{X}(t'),t')
=\frac{1}{\tau_p}\int_{-\infty}^t  \text{d}t' e^{-(t-t')/\tau_p}v(t')
=\frac{1}{\tau_p}\int_{-\infty}^{+\infty} \text{d}t' g_{\tau_p}(t-t')v(t')\;,\label{eq:CompVTauPLinear2}
\end{align}
\noindent where $g_{\tau_p}(t)=e^{-t/\tau_p}1_{t\ge 0}$.
The linear filter approximation for inertial particle velocities based on the trajectory of individual tracer particles can also be applied to the correlation functions of velocity and acceleration. To do so, $\mathcal C_v(\tau)=\langle v(t)v(t+\tau)\rangle$ defines the correlation function of the velocity of tracers where $v(t)$ is any component of the tracer velocity. Similarly, the correlation function of the inertial particle velocity $v_p(t)$ is defined as $\mathcal C_{v_p}(\tau)=\langle v_p(t)v_p(t+\tau)\rangle$ and we obtain, in the statistically stationary range,
\begin{align}\nonumber
\mathcal C_{v_p}(\tau)&=\frac{1}{\tau_p^2}(G_{\tau_p}\star \mathcal C_{v})(\tau)= \frac{1}{\tau_p^2}\int_{-\infty}^{+\infty}\text{d}t G_{\tau_p}(\tau+t)\mathcal C_{v}(t)\\
&=\frac{1}{\tau_p^2}\int_{0}^{+\infty} \text{d}t \left[G_{\tau_p}(\tau+t)+G_{\tau_p}(\tau-t)\right]\mathcal C_{v}(t)=\frac{1}{2\tau_p}\int_{0}^{+\infty} \textrm{d}t\left[e^{-|\tau+t|/\tau_p}+e^{-|\tau-t|/\tau_p}\right]\mathcal C_{v}(t)\;,
\label{eq:LinkCvpCv3}
\end{align}
where the kernel $G_{\tau_p}(t)=(g_{\tau_p}\star g_{\tau_p})(t)=\frac{\tau_p}{2}e^{-|t|/\tau_p}$ was introduced using the parity of correlation functions. Furthermore, the convolution product is defined as
\begin{equation}
 (g_1 \star g_2)(\tau)= \int_{-\infty}^{\infty} \textrm{d}t g_1(t)g_2(t-\tau)\;.
\end{equation}
The Fourier representation of Eq. (\ref{eq:LinkCvpCv3}) has been proposed for the first time by Tchen~\cite{chan2013mean}.
The determination for the inertial particle acceleration correlation function $\mathcal C_{a_p}(\tau)$ based on the filtering of the tracer acceleration $\mathcal C_{a}(\tau)$ is done in a similar fashion, resulting in
\begin{equation}
\mathcal{C}_{a_p}(\tau)=\frac{1}{2\tau_p}\int_{0}^{+\infty} \text{d}t \left[e^{-|\tau+t|/\tau_p}+e^{-|\tau-t|/\tau_p}\right]\mathcal C_{a}(t).\label{eq:LinkCapCa3}
\end{equation}
The linear filter approximation thus allows for direct assessment of velocity and acceleration correlation functions via the corresponding tracer correlation functions $\mathcal C_{v}(t)$ and $\mathcal C_{a}(t)$. The implications of this approximation will be addressed in the following section where ideal tracer, filtered inertial particles, and true inertial particles will be compared in DNS of turbulence.
\subsection{Direct numerical simulations of Lagrangian trajectories, inertial particles, and comparison to linear filter approach}
\label{sec:dns}
Twenty-one datasets containing particle trajectories for varying Stokes numbers have been generated through JHTDB (\textsc{http://turbulence.pha.jhu.edu})~\cite{yu2012studying}. The spatiotemporal evolution of the fluid velocity has been obtained by solving the Navier-Stokes equations in a periodic box with a resolution of 1024$^3$. The equation of motion of inertial particles (\ref{eq:x_p}-\ref{eq:v_p}) are solved by a second order Runge-Kutta scheme and each DNS subset of data contains $32^3$ trajectories. The relevant turbulence quantities are summarized in Tab.~\ref{tab:1}.
\begin{table}[h]
\begin{tabular}{c c c c c c c c c c}
  \centering
  Re$_{\lambda}$ & $\nu$ & $\langle \varepsilon \rangle$ & d$t$ & $\tau_{K}$ & $\tau_p$ & $T_L$ & $T_{tot}$ & $N_p$ & $N^3$ \\
  \hline
  418 & $1.85 \times 10^{-4}$ & 0.103 & 0.002 & 0.0424 & St$ \times \tau_K$ & 1.3003 & 7.692$T_L$ & $32^3$ &  $1024^3$
\end{tabular}
\caption{Characteristic parameters of the direct
numerical simulations of of inertial particles :
Taylor-Reynolds number $\textrm{Re}_{\lambda}$, kinematic viscosity $\nu$,
averaged kinetic energy dissipation rate $\langle \varepsilon \rangle $, time step of simulation ,
dissipation time $\tau_{K}= \left( \frac{\nu}{\langle \varepsilon \rangle} \right)^{1/2}$, particle response time $\tau_p$ with St $=[0,0.1,\ldots,1.9,2]$,
Lagrangian integral time $T_L$, total time of simulation $T_{tot}$, number of particles in each simulation $N_p$,
and resolution $N$ of the periodic simulation domain.}
\label{tab:1}
\end{table}

\begin{figure}[h!]
    \includegraphics[width=0.98 \textwidth]{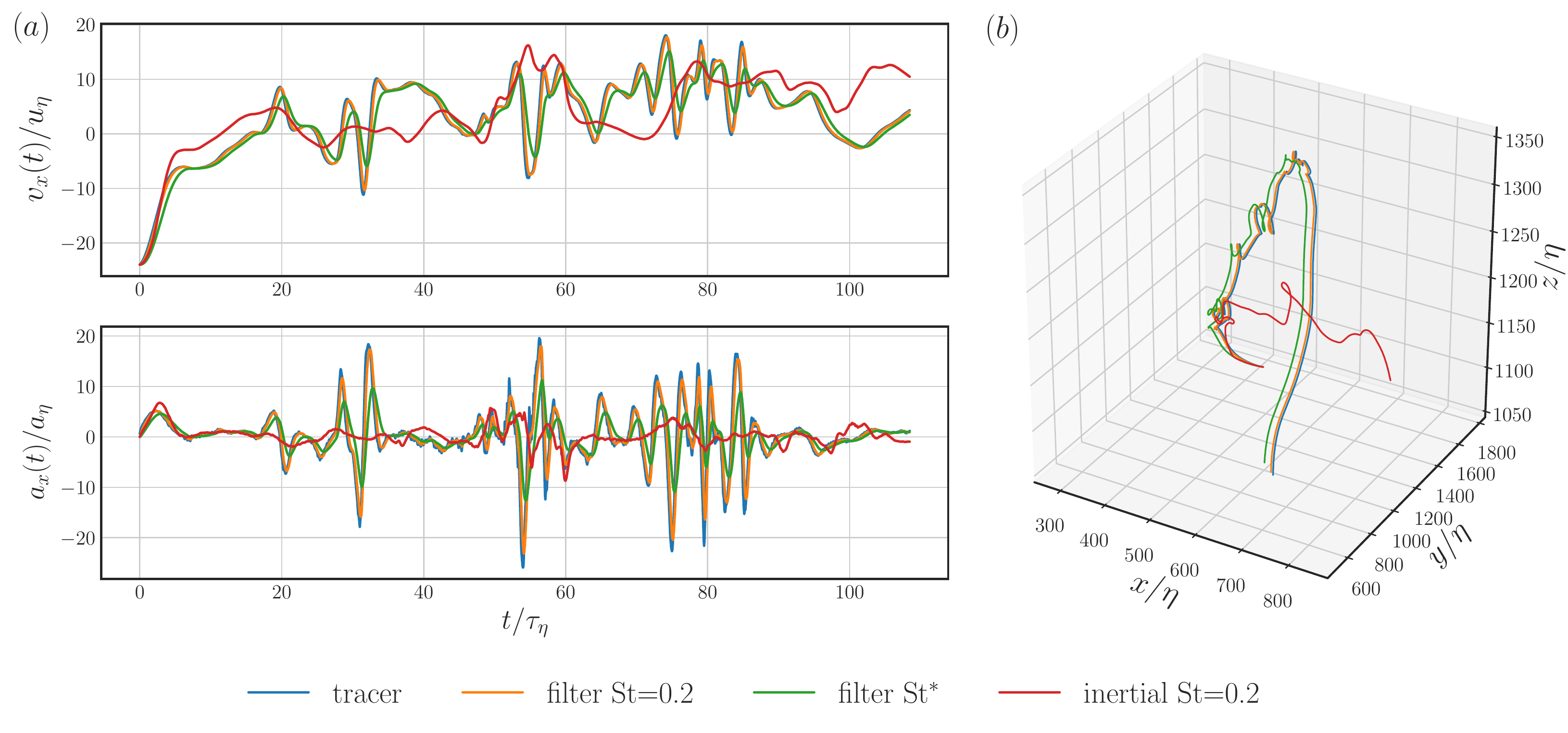}
  \caption{(a) Velocity $v_x(t)$ (upper panel) and acceleration $a_x(t)$ (lower panel) of tracers (blue), filtered tracers (orange and green) and true inertial particles for $\textrm{St}=0.2$. (b) Trajectories of tracer, filtered tracer, and inertial particles.}
  \label{fig:traj_St_0_2}
\end{figure}

The temporal evolution of particle position $\mathbf{X}_p(t)$, a velocity component $v_p(t)$, and the corresponding acceleration $a_p(t)$ for St$=0.2$ are presented in Figure~\ref{fig:traj_St_0_2} in red for a time span of $\approx 110 \tau_K$.
For comparison, Figure~\ref{fig:traj_St_0_2}(b), shows the trajectory of an ideal tracer $\mathbf{X}(\mathbf{y},t)$ (blue) starting from the same initial condition $\mathbf{X}_p(0)=\mathbf{y}$.
Whereas the tracer particle's velocity (blue) exhibits several strong oscillations, the inertial particle's velocity (red) seems not so much affected. Indeed, from Figure~\ref{fig:traj_St_0_2}(b) one can deduce that - already at such low Stokes numbers - the trajectory of the inertial particle follows a substantially different path. In the context of preferential concentration~\cite{cencini2006dynamics,Pumir_2016}, one could interpret this in terms of the inertial particle evading strong vortical flow structures. This might also be supported by the evolution of the particle's acceleration whose amplitude (and statistically speaking its variance) is significantly decreased in comparison to the tracer particle.

\begin{figure}
  \centering
  \includegraphics[width=0.49 \textwidth]{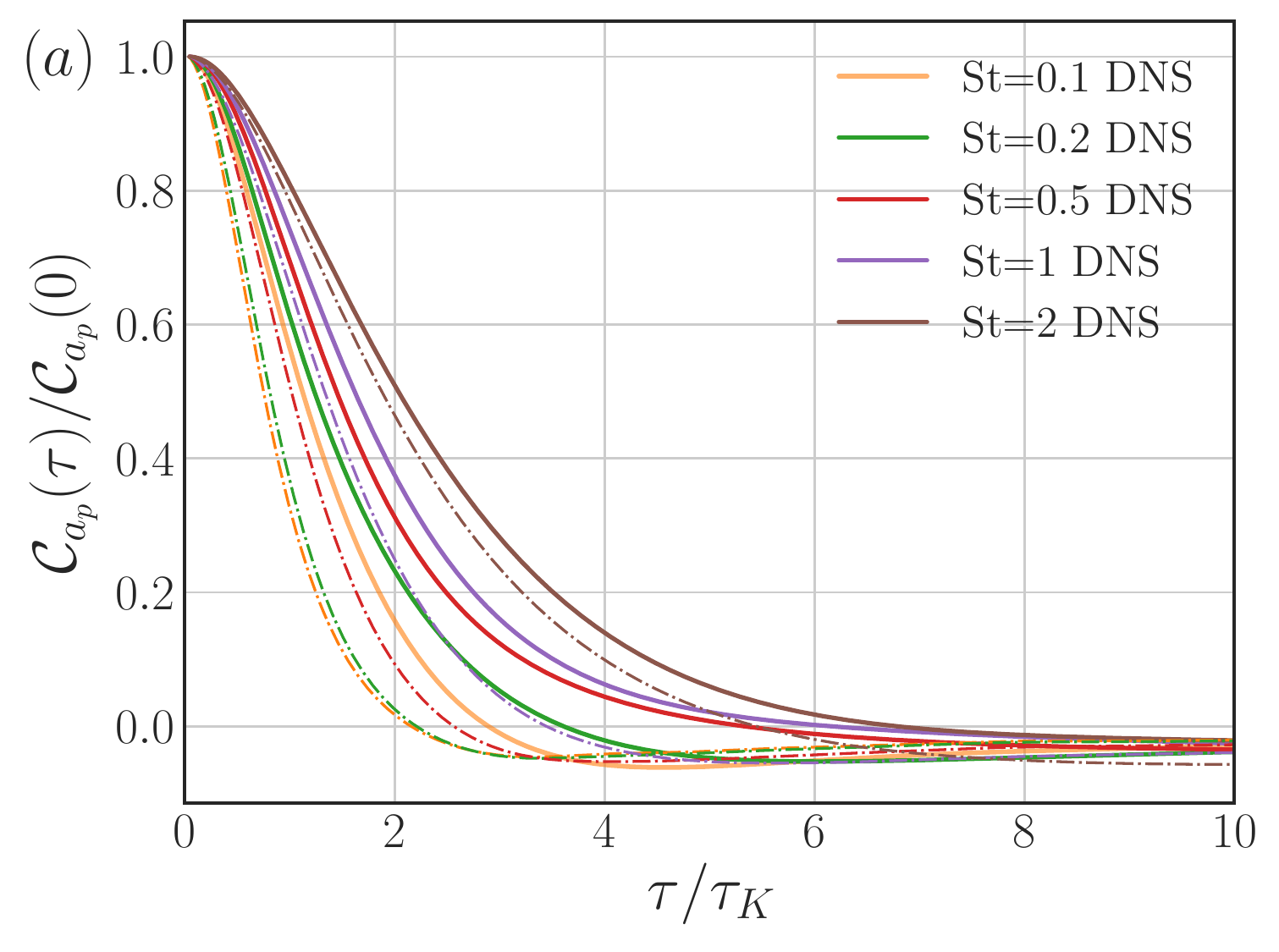}
  \includegraphics[width=0.49 \textwidth]{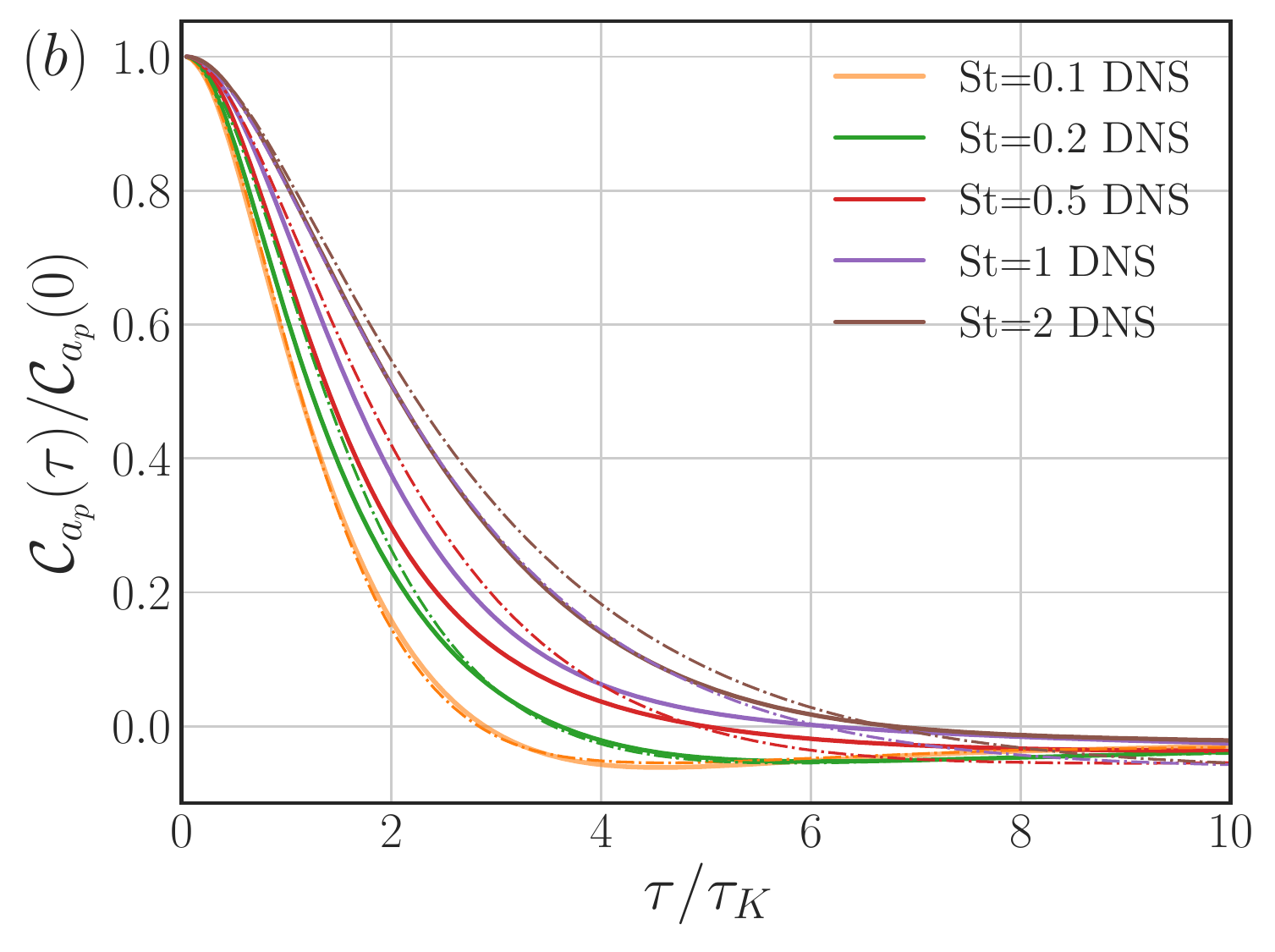}
  \caption{(a) Acceleration auto-correlation from DNS for ideal tracer (blue) and inertial particles with Stokes numbers $\textrm{St}=0.1, 0.2,0.5,1.,2.$ The dash-dotted lines correspond to the linear filter approximation (\ref{eq:x_p_filter}-\ref{eq:v_p_filter}) with $\tau_p=\textrm{St}\tau_K$ and underestimate the correlations at small $\textrm{St}$ in comparison to their DNS counterparts. The linear filter approximation approaches the DNS acceleration correlation function for larger $\textrm{St}=2$.
   (b) Same as in (a) but for a filter with an effective Stokes number $\textrm{St}^*=\tau_p^*/\tau_K$ which has been determined from the zero crossing of the acceleration correlation from DNS (dash-dotted curves and DNS curves now possess the same zero crossing). The agreement between DNS and filtered curves is more accurate than in (a) for small $\textrm{St}$.}
  \label{fig:acc_corr_filtering}
\end{figure}

The linear filter approximation from Eqs. (\ref{eq:x_p_filter}-\ref{eq:v_p_filter}) with $\tau_p=0.2 \tau_K$  (St$=0.2$, orange curve) remains very close to its determining tracer trajectory (blue). Especially the temporal evolution of the acceleration suggests that the linear filter (orange) overestimates the acceleration variance in comparison to the ordinary inertial particle.
Cencini $et\,al.$~\cite{cencini2006dynamics} applied the same technique to compare root-mean-squared accelerations and observed large discrepancies between filtered and true $a_{rms}$ at low $\textrm{St}$. Nonetheless, for larger Stokes numbers, true and filtered $a_{rms}$-values seemed to approach one another. Hence, this study suggests that non-trivial effects of the preferential concentration, when inertia is introduced to the particles, creates the gap between true and filtered results for $\textrm{St}<$1. On that account, Figure~\ref{fig:acc_corr_filtering}(a) depicts the acceleration auto-correlation function which has been obtained by averaging over all
$N_p=32^3$ particle trajectories. Here, the dash-dotted curves correspond to the linear filter approximation (\ref{eq:LinkCapCa3}) for $\tau_p=\text{St}\tau_K$ with St$=[0.1,0.2,0.5,1,2]$ from left to right. Thereby, the fluid velocity correlation function $\mathcal{C}_a(\tau)$ was integrated over the entire time range of the
simulation $T$ (see also Tab.~\ref{tab:1}). In agreement with the findings by Cencini $et\,al.$~\cite{cencini2006dynamics}, the modeling breaks down initially, drastically underestimating the correlation of the
particle for $\textrm{St} \le $1, but recovers to generate meaningful statistics for $\textrm{St}$=2. Especially, the regime of low Stokes numbers, e.g., for $\textrm{St}$=0.1, suggests that the filter approximation (first dash-dotted curve from left) is nearly identical to the ideal tracer $\textrm{St=0}$ (not shown in plot) and decorrelates much faster than the true DNS for $\textrm{St}$=0.1 (orange curve).

\begin{figure}[h!]
    \includegraphics[width=0.55 \textwidth]{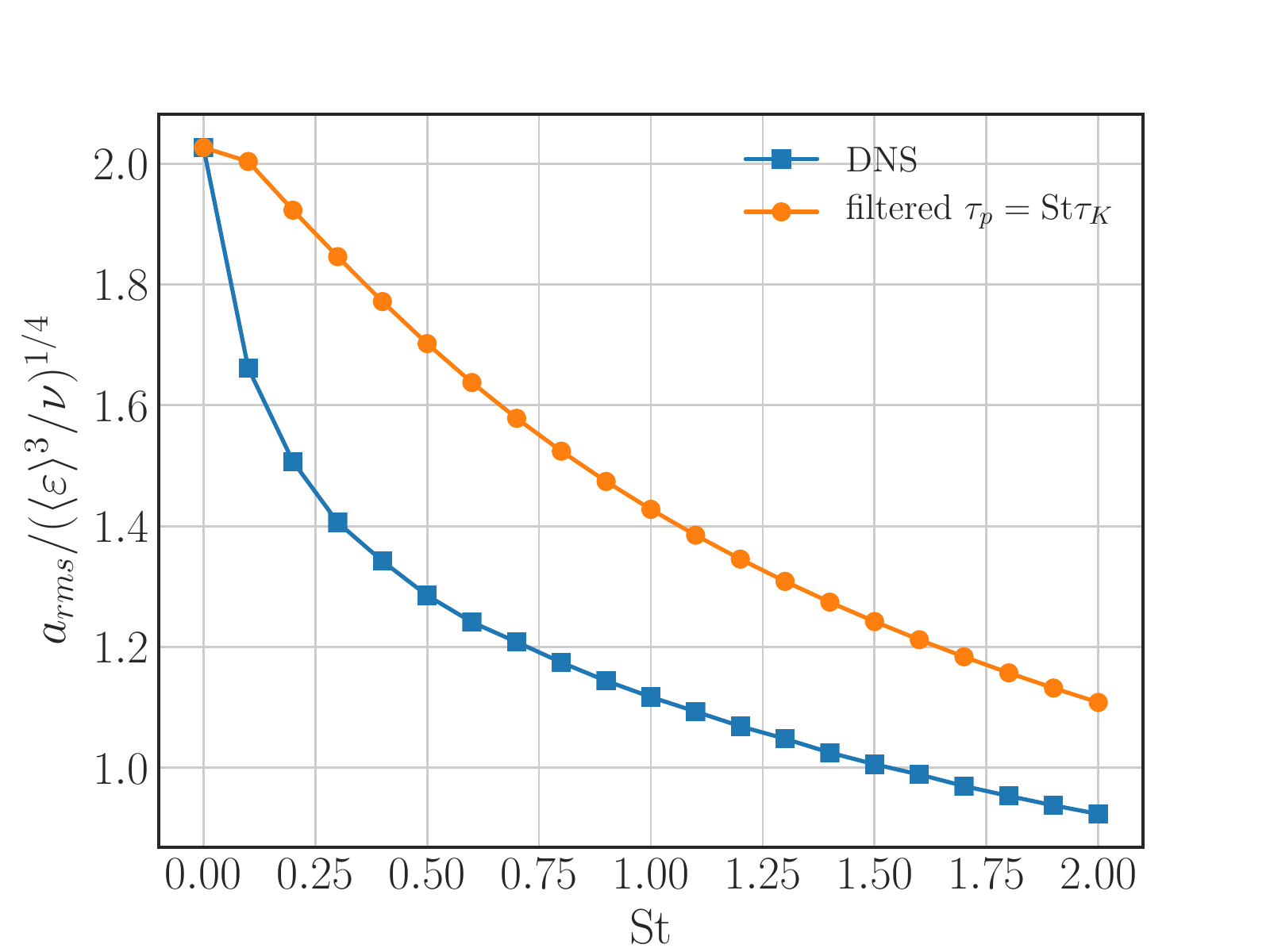}
  \caption{Root mean square values of acceleration $a_{rms}$ as a function of the Stokes number for inertial particles in DNS (blue) and filtered tracer particles (orange) according to Eq. (\ref{eq:v_p_filter}). The $a_{rms}$-values from DNS exhibit a rapid decrease for small but finite $\textrm{St}$ whereas the linear filter approximation decays much slower.}
  \label{fig:a_rms_DNS_filter}
\end{figure}

These profound changes between inertial particle and tracers (or filtered tracers) acceleration properties at low Stokes also manifest themselves in the root mean square of acceleration itself. Figure~\ref{fig:a_rms_DNS_filter} depicts the root mean square values of acceleration $a_{rms}$ as a function of $\textrm{St}$ for inertial particles (blue) and filtered tracer particles (orange) according to Eq. (\ref{eq:CompVTauPLinear2}). The
$a_{rms}$-values of inertial particles decreases much faster than their filtered counterparts. These strong discrepancies were interpreted by Cencini $et\,al.$~\cite{cencini2006dynamics} in terms of inertial particles which preferentially sample regions of low turbulence intensity (or depleted vorticity regions) whereas the filtered tracer particles are still impacted by strong acceleration events of tracer trajectories trapped in vortical structures. In other words, by restricting itself to individual tracer trajectories $\mathbf{X}(\mathbf{y},t)$, the filtering approach bears no information on the spatial organization of the surrounding fluid velocity field, which apparently is crucial for a better understanding of the dynamics of inertial particles.

In the following, we suggest a modification of the linear filter which is in better agreement with DNS but at the same time still is solely based on the Lagrangian velocity $\mathbf{v}(\mathbf{y},t)$ of an individual tracer. To this end, we introduce an effective particle response time $\tau_p^*$ in the linear filter approximation (\ref{eq:v_p_filter}) which does not necessarily obey the usual relation $\tau_p = \text{St}\tau_K$ but henceforward is considered as a free model parameter. In the spirit of recent stochastic models for Lagrangian tracers~\cite{Viggiano_2020}, which identified the zero crossing of the acceleration correlation as a crucial ingredient for model calibration, we proceed in a similar way: First, we determine the zero crossing $\tau_0(\textrm{St})$ of the acceleration auto-correlation functions
from DNS in Figure~\ref{fig:acc_corr_filtering}. Subsequently, with the help of a suitable roots-finding algorithm, we determine the effective particle
response time $\tau_p^*(\textrm{St})$ in Eq. (\ref{eq:LinkCapCa3}) in such a way that $\mathcal{C}_{a_p}(\tau_0(\textrm{St}))=0$.

Figure~\ref{fig:acc_corr_filtering}(b) depicts the acceleration auto-correlation functions (dash-dotted curves) after this calibration. Hence, by virtue of an effective particle response time $\tau_p^*$ based on matching zero crossing of the acceleration correlation functions, a better agreement with DNS has been achieved, particularly for low Stokes numbers. The corresponding filtered trajectories with St$^*=\tau_p^*/\tau_K$ are also included in Figure~\ref{fig:traj_St_0_2} and correspond to the green curves. Due to the increased damping ($\tau_p^*>\tau_p$)
in the filter, $\tau_p^*$-filtered accelerations (green) are closer to their DNS counterparts (red). In conclusion, the introduction of an effective particle response time $\tau_p^*$ based on the zero crossing of the acceleration correlation function reinstates the effects of preferential sampling of inertial particles to some extend. This notion will be further assessed in the following section by example of stochastic models for inertial particle statistics.
\section{Process for inertial particles subject to Gaussian infinitely differentiable fluid velocity}
\label{sec:stoch_model}
This section is devoted to a generalization of a recently developed model for the Lagrangian velocity~\cite{Viggiano_2020} which will be extended to take into account finite particle inertia. The simplest model for the Lagrangian velocity $v(t)$ consists of an Ornstein-Uhlenbeck process which obeys the following Langevin equation
\begin{equation}
  \textrm{d} v(t) =- \frac{1}{T}  v(t) \textrm{d}t + \sqrt{\frac{2\sigma^2}{T}} \textrm{d}W(t)\;,
\end{equation}
where $\textrm{d}W(t)$ denote the increments of a Wiener process $W(t)$ with zero mean and variance $\textrm{d}t$, and $\sigma^2$ denotes the variance of the process $\langle v^2 \rangle$. Furthermore, $T$ is a large turbulence time scale. In this framework, velocity is non-differentiable and thus acceleration only has a meaning in a distributional sense, which is at odds with experimental and numerical observations. This model has been extended by Sawford~\cite{sawford1991reynolds} in order to allow for the differentiability of the velocity, however, the acceleration of
the process remains non-differentiable itself. Therefore, a multilayered Ornstein-Uhlenbeck process has been proposed where the velocity itself is infinitely differentiable when the number of layers tends to infinity~\cite{Viggiano_2020}. In making use of the linear filter approximation discussed in the previous section, the latter model can readily be extended in order to allow for the effects of finite particle inertia.
Following the approach of \cite{Viggiano_2020}, we consider the system of coupled stochastic differential equations
\begin{align}
 \dot v_{p}  =& \frac{1}{\tau_p^*}\left(v- v_{p}\right)\;, \label{eq:myvp} \\
 \dot v  =& - \frac{1}{T}  v +f_{\tau_\eta}\;, \label{eq:myv}
\end{align}
where $\tau_p^*$ denotes the particle response time of the model and $\tau_\eta$ is a small-scale turbulence time scale of the order of the Kolmogorov dissipative time scale $\tau_K$ defined in the caption of table \ref{tab:1}. The forcing term $f_{\tau_\eta}$ is a zero-average Gaussian random force, which is fully determined by its covariance in the statistically stationary regime. Following the procedure outlined in~\cite{Viggiano_2020}, the covariance in both, physical and spectral space, reads
 \begin{align}\label{eq:CofFTaueta}
 \mathcal C_{f_{\tau_\eta}}(\tau) \equiv \langle f_{\tau_\eta}(t)f_{\tau_\eta}(t+\tau)\rangle = \frac{\sigma^2e^{-\tau_\eta^2/T^2}}{\sqrt{\pi}T\tau_\eta \erfc\left(\tau_\eta/T\right)}e^{-\frac{\tau^2}{4\tau_\eta^2}} = \frac{2\sigma^2e^{-\tau_\eta^2/T^2}}{T \erfc\left(\tau_\eta/T\right)} \int \textrm{d}\omega\; e^{2i\pi \omega\tau}e^{-4\pi^2\tau_\eta^2\omega^2}\;,
\end{align}
where we have introduced  $\erfc(x)=1-\erf(x)$ as the complementary of the error function $\erf(x)$. As we can see from the structure of its covariance (\ref{eq:CofFTaueta}), the Gaussian forcing term $f_{\tau_\eta}$  is correlated over the dissipative time scale $\tau_\eta$, the correlation function being itself a Gaussian function. Moreover, this covariance structure shows that the random process $f_{\tau_\eta}(t)$ is infinitely differentiable, which is a direct consequence of the smoothness of its Gaussian shape. As developed in Ref.~\cite{Viggiano_2020}, the remaining parameter $\sigma^2$ entering Eq. \ref{eq:CofFTaueta} corresponds to the variance of the Lagrangian velocity $v$, that is the variance of the solution of Eq. (\ref{eq:myv}) in the statistically steady range
\begin{align}\label{eq:DefSigma}
\sigma^2=\langle v^2\rangle.
\end{align}
At this stage, causality of the forcing term $f_{\tau_\eta}$ is not obvious, and cannot be deduced from the covariance $\mathcal C_{f_{\tau_\eta}}(\tau)$ (Eq. \ref{eq:CofFTaueta}). It is indeed shown in Ref.~\cite{Viggiano_2020} that this forcing term can be obtained as an infinite sequence of causal equations, i.e., an infinite number of embedded Ornstein-Uhlenbeck processes, making the asymptotic process $f_{\tau_\eta}$ itself causal.
Finally, the covariance function of the particle velocity $v_p$, defined as successive linear operations (Eqs. \ref{eq:myvp} and \ref{eq:myv}) on the forcing term $f_{\tau_\eta}$, can be obtained as
 \begin{equation}
\mathcal C_{v_p}(\tau)\equiv   \langle v_{p}(t)v_{p}(t+\tau)\rangle  = \frac{2\sigma^2e^{-\tau_\eta^2/T^2}}{T\erfc\left(\tau_\eta/T\right)}\int \textrm{d}\omega\; e^{2i\pi \omega\tau}\frac{1}{1+4\pi^2(\tau_p^*)^2
\omega^2}
\frac{T^2}{1+4\pi^2T^2\omega^2}e^{-4\pi^2\tau_\eta^2\omega^2}\;,
\label{eq:CorrVelLimFour2}
\end{equation}
which can be calculated explicitly in physical space according to
\begin{align} \nonumber
\mathcal C_{v_p}(\tau)=& \frac{\sigma^2 T}{2(T^2-(\tau_p^*)^2)\erfc(\tau_\eta/T)}\Bigg \{ T e^{-|\tau|/T} \left[\erfc\left(\frac{\tau_\eta}{T}-\frac{|\tau|}{2\tau_\eta}\right)+e^{2|\tau|/T}
\erfc\left(\frac{|\tau|}{2\tau_\eta}+\frac{\tau_\eta}{T}\right)\right] \\
&-\tau_p^* e^{-|\tau|/\tau_p^*} e^{\frac{\tau_\eta^2}{(\tau_p^*)^2}-\frac{\tau_\eta^2}{T^2}}
\left[ \erfc\left(\frac{\tau_\eta}{\tau_p^*}-\frac{|\tau|}{2\tau_\eta}\right)+e^{2|\tau|/\tau_p^*}
\erfc\left(\frac{|\tau|}{2\tau_\eta}+\frac{\tau_\eta}{\tau_p^*}\right)\right]
 \Bigg \}.
\label{eq:ExplicitCorrVLim}
\end{align}
Herein, the dependence of the variance of the particle velocity on $\tau_p^*$ can be determined. Setting $\tau=0$ in equation (\ref{eq:ExplicitCorrVLim}) yields
\begin{equation}
  \sigma_{v_p}^2=
  \frac{\sigma^2 T^2}{T^2-(\tau_p^*)^2}\left[1- \frac{\tau_p^*}{T} e^{\frac{\tau_\eta^2}{(\tau_p^*)^2}-\frac{\tau_\eta^2}{T^2}} \frac{\erfc\left(  \frac{\tau_\eta}{\tau_p^*}\right)}{\erfc\left(\frac{\tau_\eta}
  {T}\right)}
  \right]\;.\label{eq:tau_pstar}
\end{equation}
Hence, in the limit $\tau_p^* \rightarrow 0$, the variance of the Lagrangian model, $\lim_{\tau_p^*\rightarrow 0} \sigma_{v_p}^2=\sigma^2$, is recovered~\cite{Viggiano_2020}. Furthermore, the prediction of the ``single-layered'' Tchen-Hinze theory (\ref{eq:tchen}) can be recovered by discarding the angular brackets in Eq. (\ref{eq:tau_pstar}).\\
The acceleration correlation function can be calculated from Eq. (\ref{eq:ExplicitCorrVLim}) according to
\begin{equation}\label{eq:ca_MF}
\mathcal C_{a_p}(\tau) = -\frac{\textrm{d}^2\mathcal C_{v_p}(\tau)}{\textrm{d}\tau^2},
\end{equation}
which yields
\begin{align} \nonumber
\mathcal C_{a_p}(\tau)=& \frac{\sigma^2T}{2(T^2-(\tau_p^*)^2)\erfc(\tau_\eta/T)}\Bigg \{ \frac{1}{T}e^{-|\tau|/T} \left[\frac{2T}{\sqrt{\pi}\tau_\eta} e^{-\left(\frac{\tau_\eta}{T}
-\frac{|\tau|}{2\tau_\eta}\right)^2}- \erfc\left(\frac{\tau_\eta}{T}-\frac{|\tau|}{2\tau_\eta}\right)-e^{2|\tau|/T}
\erfc\left(\frac{|\tau|}{2\tau_\eta}+\frac{\tau_\eta}{T}\right)\right] \\
&-\frac{1}{\tau_p^*} e^{-|\tau|/\tau_p^*} e^{\frac{\tau_\eta^2}{(\tau_p^*)^2}-\frac{\tau_\eta^2}{T^2}}
\left[\frac{2 \tau_p^*}{\sqrt{\pi}\tau_\eta} e^{-\left(\frac{\tau_\eta}{\tau_p^*}
-\frac{|\tau|}{2\tau_\eta}\right)^2}- \erfc\left(\frac{\tau_\eta}{\tau_p^*}-\frac{|\tau|}{2\tau_\eta}\right)-e^{2|\tau|/\tau_p^*}
\erfc\left(\frac{|\tau|}{2\tau_\eta}+\frac{\tau_\eta}{\tau_p^*}\right)\right]
 \Bigg \}.
\label{eq:ExplicitCorrALim}
\end{align}
The variance of the acceleration can be determined according to
\begin{equation}
  \sigma_{a_p}^2 = \frac{\sigma^2}{T^2-(\tau_p^*)^2}\left [\frac{T}{\tau_p^*}
 \frac{e^{\frac{\tau_\eta^2}{(\tau_p^*)^2}}}{e^{\frac{\tau_\eta^2}{T^2}}}\frac{\erfc\left(\frac{\tau_\eta}{\tau_p^*}\right)}{\erfc\left(\frac{\tau_\eta}{T}\right)}-1
 \right ]\;.
\end{equation}
In the limit of $\tau_p^* \rightarrow 0$, the reduced tracer model discussed in~\cite{Viggiano_2020} is recovered and we get
\begin{equation}
  \lim_{\tau_p^*\rightarrow 0} \sigma_{a_p}^2 = \frac{\sigma^2}{T^2}\left [\frac{T}{\sqrt{\pi}\tau_\eta}
 \frac{e^{-\frac{\tau_\eta^2}{T^2}}}{\erfc\left(\frac{\tau_\eta}{T}\right)}-1
 \right ]\;.
\end{equation}
So far, the proposed model possesses Gaussian properties, i.e., the particle velocity statistics is fully determined by the correlation function (\ref{eq:ExplicitCorrVLim}). The inclusion of non-Gaussian properties into the stochastic process can be achieved in the same manner discussed in~\cite{Viggiano_2020} and will be the subject of future work. In the present model, the fluid properties enter through the turnover time $T$ and the small time scale $\tau_\eta$, which can be parameterized in order to match certain characteristics of Lagrangian tracers. In~\cite{Viggiano_2020}, for instance, $T$ and $\tau_\eta$ were determined in order to match the integral Lagrangian time scale
\begin{equation}
    T_L = \int_0^\infty \textrm{d}\tau \frac{\mathcal C_v(\tau)}{\mathcal C_v(0)}\;,
    \label{eq:T_L}
\end{equation}
and the zero crossing of the empirically determined acceleration correlation function $\mathcal C_a(\tau)$. In addition to this parameterization, here, we also have to determine the particle response time $\tau_p^*$. In the spirit of the effective particle response time of the linear filter approximation discussed in Section~\ref{sec:lin_filt}, $\tau_p^*$ will be determined in order to match the zero crossing of the acceleration correlation function for different Stokes numbers. However, before we explicitly carry out the model calibration, we want to discuss the linear filter approximation in the context of the Batchelor model for Lagrangian tracers in turbulence. The latter modeling technique operates directly at the level of the second order structure function $\mathcal S_2(\tau) = 2\sigma^2 - 2C_v(\tau)$ and differentiability of the Lagrangian velocity is ensured by introducing the correct dissipation range behavior at small $\tau$. By contrast to the present stochastic model, which was based on a Langevin equation (\ref{eq:myvp}) at its basic level, the Batchelor model will not be discussed in form of a stochastic process, but rather in the context of the so-called multifractal formalism which operates directly on the structure function level~\cite{arneodo2008universal}.

\section{Inertial particle statistics from Batchelor's model for Lagrangian tracers}
\label{sec:batchelor}
In the seminal work of Batchelor~\cite{Batchelor1951} an interpolation between dissipation and inertial range was proposed in the Eulerian frame of reference. The interpolation was capable to reproduce the inertial range behavior of the second order structure function $\langle (\delta_r u) \rangle \sim r^{2/3}$ and, at small $r$, the dissipation range prediction $\langle (\delta_r u) \rangle \sim r^{2}$ by a simple polynomial interpolation~\cite{meneveau1996transition}. Similarly, an interpolation for the Lagrangian second order structure function
$\left \langle (\delta_\tau v)^2 \right \rangle$, which covers both inertial $\left \langle (\delta_\tau v)^2 \right \rangle \sim \tau$ and dissipation range $\tau^2$, can be deduced~\cite{chevillard2003lagrangian,chevillard2012phenomenological,arneodo2008universal,Benzi2010}. In its simplest form, the second order Lagrangian structure function reads
\begin{equation}
  S_2(\tau)= \left \langle (\delta_\tau v)^2 \right \rangle =2 \sigma^2 \frac{\frac{\tau}{T}}{\left [1+ \left(\frac{\tau}{\tau_\eta} \right)^{-\delta}\right]^{\frac{1}{\delta}}}\;,
  \label{eq:batch_s2}
\end{equation}
where $\delta$ determines the transition between dissipation and inertial range. Within the multifractal formalism, which proposes a more complete modeling of the differential action of viscosity not only up to second order statistics as it is done here, but also for higher order structure functions, Lagrangian statistics seem to be well reproduced by choosing $\delta=4$. Nevertheless, in what follows, where we completely neglect the implications of intermittent corrections on the width and shape of the so-called intermediate dissipative range, we will restrict ourselves to the case $\delta=2$, as initially chosen by Batchelor to reproduce the behavior of the second order structure function of Eulerian velocity, which leads to an acceptable agreement with DNS data, as it will be shown in the following Section. Furthermore, because the proposed parametrization for $S_2$ (equation \ref{eq:batch_s2} is understood as a model for scales $\tau\ll T$, we need to come up with a proper cutoff at large scales if we intend to use such a form for $\tau\gg T$). Following the proposition of Ref. \cite{Viggiano_2020}, we include in this simple picture an additional exponential decrease over the characteristic large time scale $T$, and consider the alternative form
\begin{equation}
S_2(\tau)=2 \sigma^2 \frac{1-e^{-\frac{\tau}{T}}}{\left [1+ \left(\frac{\tau}{\tau_\eta} \right)^{-\delta}\right]^{\frac{1}{\delta}}}\;,
  \label{eq:batch_s2_new}
\end{equation}
which entails a finite Lagrangian integral time scale. For our case, i.e., $\delta=2$, this time scale can be calculated explicitly
\begin{equation}
  T_L =\int_0^{\infty} \textrm{d}\tau \frac{\mathcal C_v(\tau)}{\mathcal C_v(0)}= \tau_\eta \left[ 1-\frac{\pi}{2} Y_1\left(\frac{\tau_\eta}{T}\right)
-\frac{\pi}{2} H_{-1}\left(\frac{\tau_\eta}{T}\right) \right]\;,
\label{eq:T_L_batch}
\end{equation}
where $Y_n(z)$ denotes the Bessel function of the second kind and $H_{n}(z)$ the Struve function, as it is provided by a symbolic calculation software. In this framework,  the acceleration correlation function for the tracer particle can be calculated from the second order structure function (\ref{eq:batch_s2}) by $\mathcal C_{a}(\tau) = \frac{1}{2}\frac{\textrm{d}^2\mathcal S_{2}(\tau)}{\textrm{d}\tau^2}$.
%

Finite particle inertia can again be included in using the linear filter approximation discussed in Section~\ref{sec:lin_filt}. To this end, Eq. (\ref{eq:LinkCvpCv3}) has to be evaluated from the Lagrangian velocity correlation $\mathcal C_v(\tau) = \sigma^2 - \frac{S_2(\tau)}{2}$. We could not obtain a closed expression for the integral and thus have restricted ourselves to a numerical evaluation, which will be further elaborated upon in the next section.

\section{Comparison to DNS}
\label{sec:DNScompare}
In this section, we will apply the presented modeling techniques to the DNS discussed in Section~\ref{sec:dns}. In order to connect these approaches to the simulated data, parameters of the DNS must be defined to properly calibrate the models, namely the integral length scale, $T$, the dissipative scale, $\tau_{\eta}$, and the effective particle response time $\tau_p^*$. Detailed explanations of the calibration technique can be found in \cite{Viggiano_2020}. In summary, two conditions have to be imposed for the Lagrangian stochastic model ($\tau_p^*=0$), namely the matching of the Lagrangian time scale of the stochastic model (\ref{eq:T_L}) to $T_L$ of the DNS as well as the matching of the  zero crossing of the acceleration correlation function (\ref{eq:ExplicitCorrALim}) to the zero crossing from the DNS data. This calibration suggests the values $(\tau_\eta/\tau_K, T/T_L)= (0.5759, 0.9791)$.
For the Batchelor model, the same calibration has to be carried out, where now the Lagrangian integral time scale is given by Eq. (\ref{eq:T_L_batch}). Furthermore,
by matching the zero crossing, we obtain the values $(\tau_\eta/\tau_K, T/T_L)= (1.7956, 0.9941)$. It can be noted that $\tau_\eta$ from the Batchelor model is
larger than $\tau_K$ whereas the stochastic model exhibits a $\tau_\eta$ smaller than $\tau_K$ after calibration.

With the inclusion of $\textrm{St}$, an additional free parameter of the models is available for calibration, the effective particle response time $\tau_p^*$.
As discussed, the acceleration correlation function for the stochastic Gaussian process is given by Eq. (\ref{eq:ExplicitCorrALim}), from which $\tau_p^*$ can be extracted based on the zero crossing of the DNS data for each Stokes number.
In a similar fashion, the new model parameter $\tau_p^*$ is obtained from the linear filtering of the acceleration correlation function of the Batchelor model derived from Eq. (\ref{eq:batch_s2_new}).

\begin{figure}
  \centering
  \includegraphics[width=0.49 \textwidth]{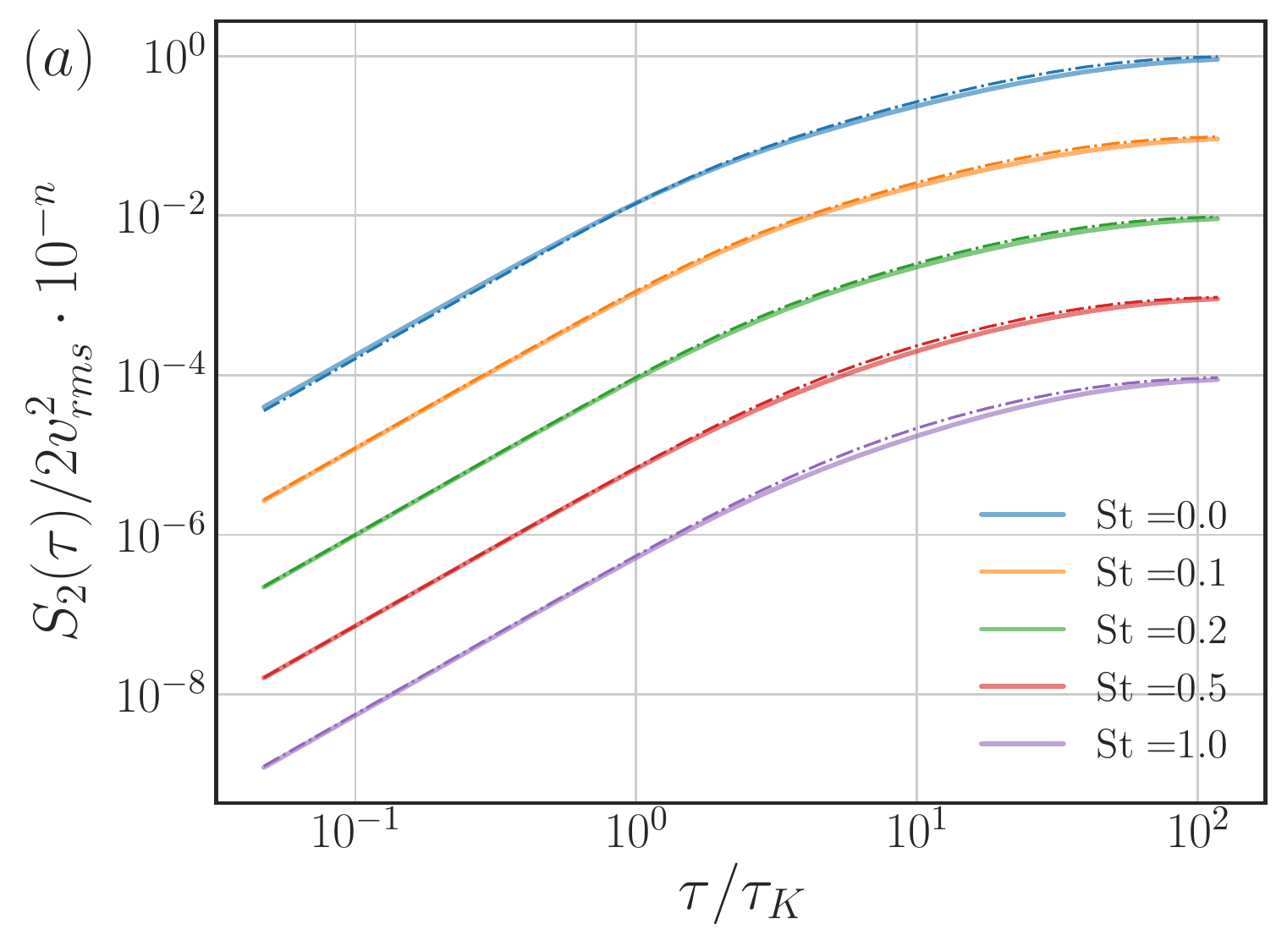}
  \includegraphics[width=0.49 \textwidth]{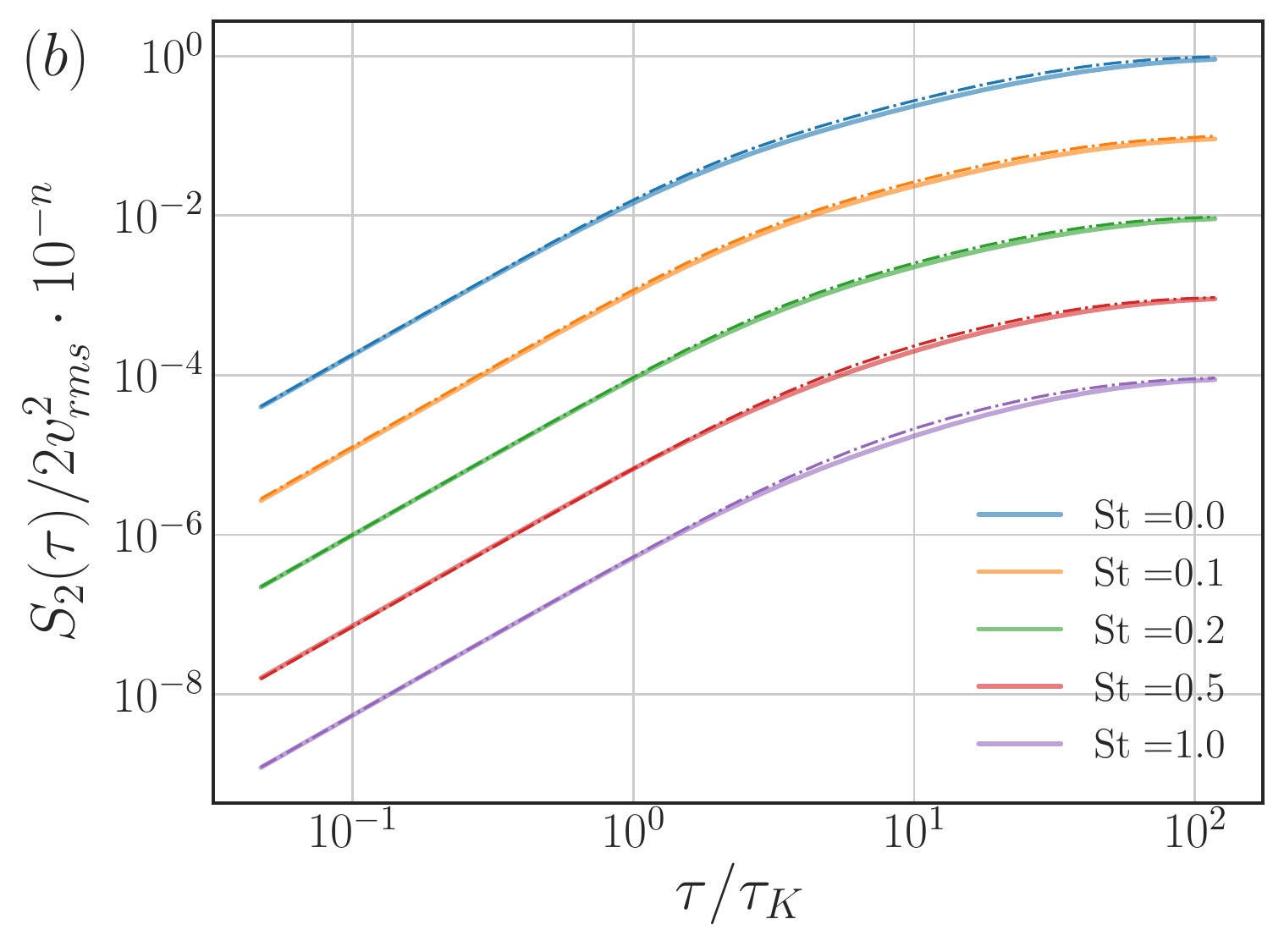}
  \caption{Comparison of the second-order structure function to DNS from (a) the stochastic model (dash-dotted lines) and (b) the Batchelor model for 5 different $\textrm{St}$ parameters in between 0 and 1. The structure functions have been shifted vertically by multiplication of $10^{-n}$ for $n=0-4$ from top to bottom.}
  \label{fig:s2}
\end{figure}

Figures~\ref{fig:s2}(a) and~\ref{fig:s2}(b) depict the comparison of second order structure function $S_2(\tau)= \left \langle (\delta_\tau v_{p})^2 \right \rangle$  stochastic and Batchelor model, respectively (dash-dotted curves), to DNS for $\textrm{St}$=0-1 (solid curves). For the tracers, $\textrm{St}$=0, the original stochastic model~\cite{Viggiano_2020} and
Batchelor model are implemented.

\begin{figure}
  \centering
  \includegraphics[width=0.49 \textwidth]{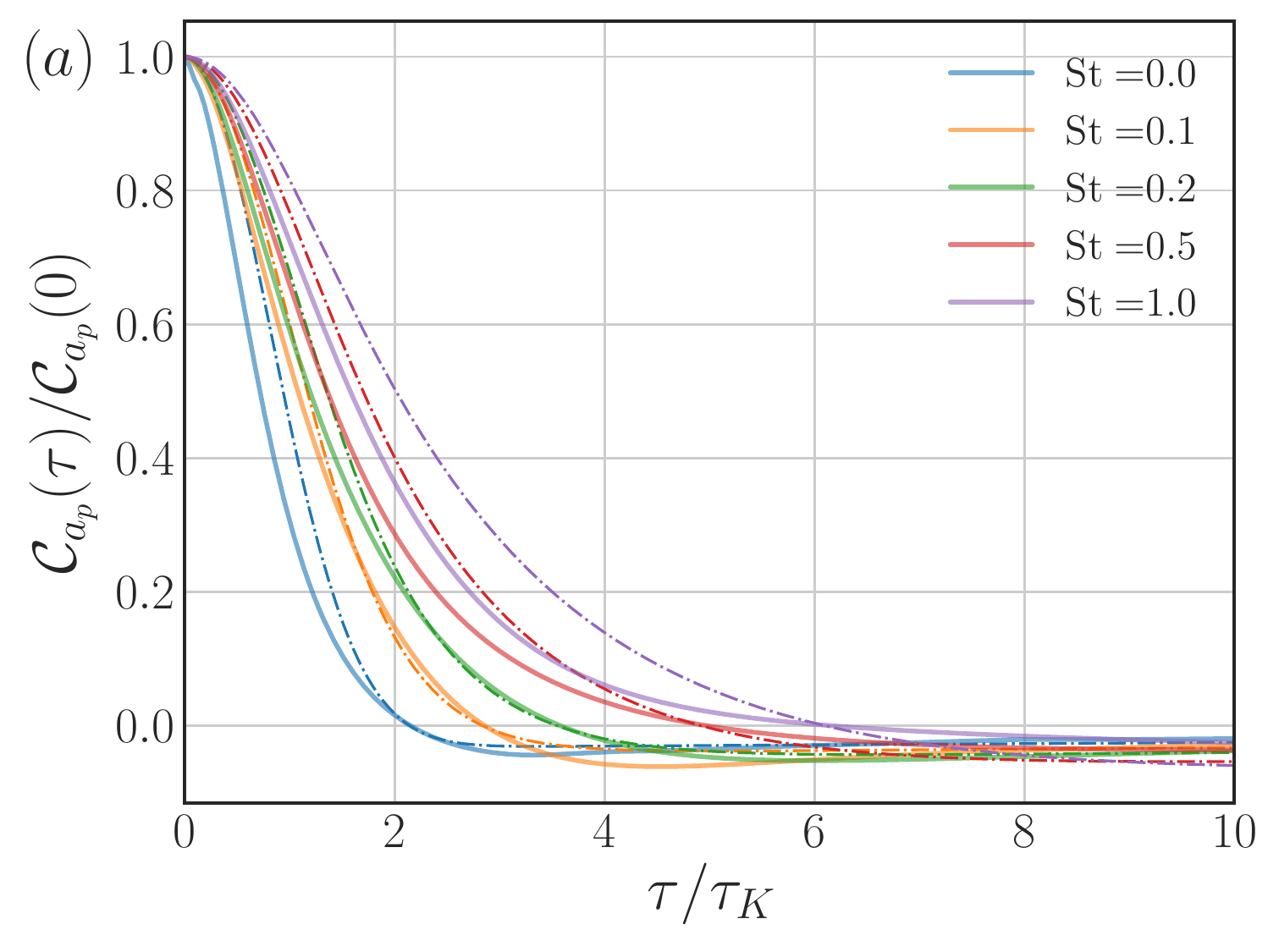}
  \includegraphics[width=0.49 \textwidth]{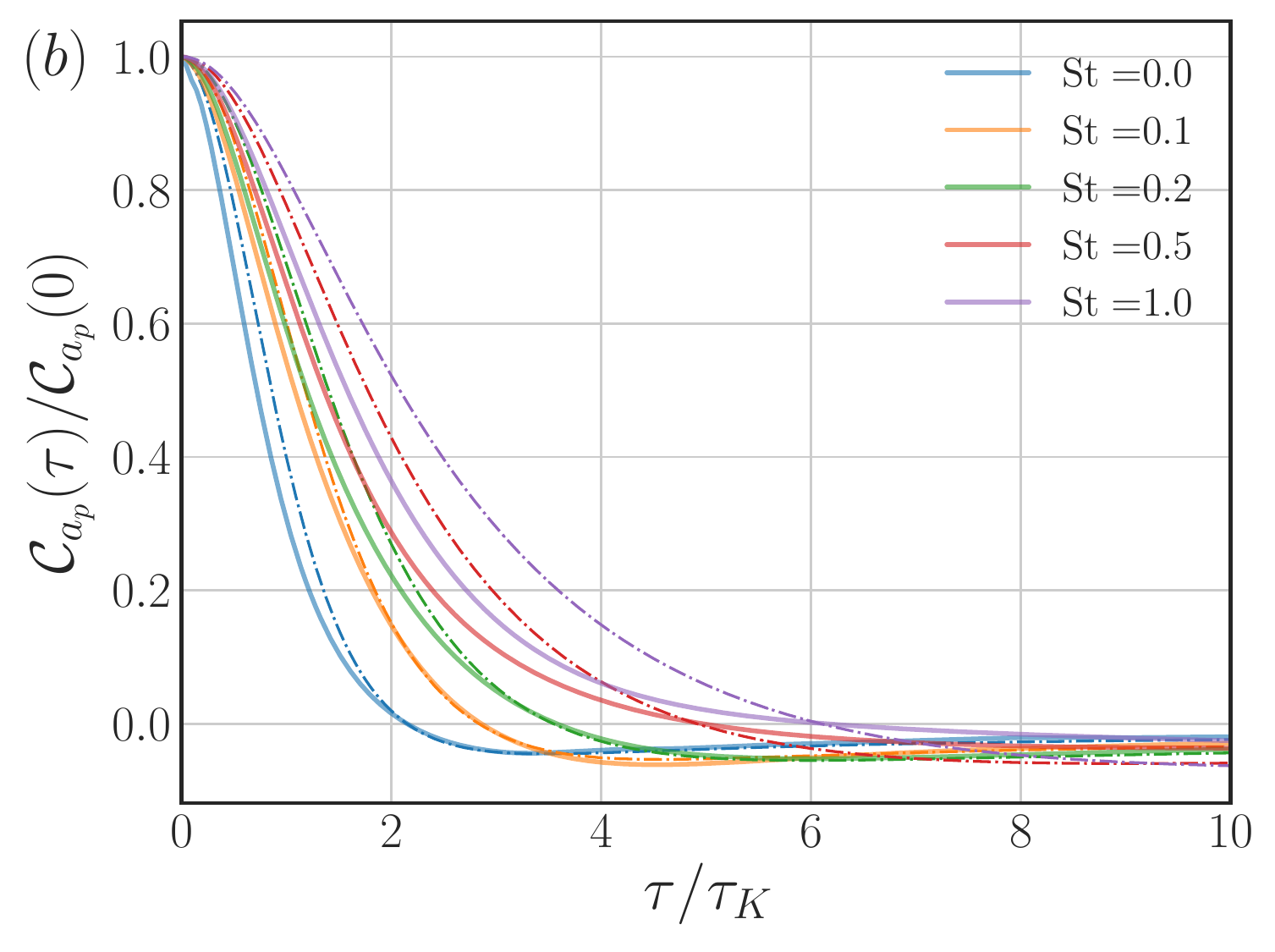}
  \caption{Comparison of acceleration auto-correlation function from (a) the stochastic model (dash-dotted lines) and (b) the Batchelor model (dash-dotted lines) to DNS for 5 different $\textrm{St}$ parameters in between 0 and 1.}
  \label{fig:acc_corr}
\end{figure}

For $\textrm{St}>$0, linear filtering of the velocity correlation function for tracer particles agrees well with $S_2(\tau)$ from DNS for both models. The stochastic approach, Figure~\ref{fig:s2}(a) shows agreement between the model and DNS at small scales and deviates slightly as the time lag $\tau$
increases. Notably, slight deviations appear in the inertial range and might be attributed to the Gaussianity of the stochastic model, which thus neglects intermittency corrections. The application of the our filtering technique to the statistics of the Batchelor model, Figure~\ref{fig:s2}(b), shows similar tendencies. At small scales the model coincides with the DNS profiles for the given
$\textrm{St}$ presented. At $\tau/\tau_K>$1, again a deviation occurs where the model begins to overestimate the structure function of its corresponding DNS curve, the near-dissipative range seems to extend further than the one present in the simulated data. These deviations slightly increase with $\textrm{St}$.
The application of the models to the acceleration correlation function is presented in Figure~\ref{fig:acc_corr}. Here, the discrepancy between modeled correlations and those obtained from the simulated data can be observed at all scales. The stochastic modeling of the inertial particle correlations is presented in Figure~\ref{fig:acc_corr}(a) for the considered $\textrm{St}$ range.  The model over-predicts correlations in the dissipative range. For increasing Stokes number, linear filtering of the model improves the ability to accurately describe small scale correlations and still the large scale variations between the two curves are minimal, for example for $\textrm{St}$=0.2. As $\textrm{St}$ increases further, quickly the filtered model and the DNS show increased discrepancies between the profiles, as the decorrelation of the acceleration occurs more rapidly than the model predicts. Filtering of the Batchelor model, Figure~\ref{fig:acc_corr}(b), shows similar tendencies of the results of the stochastic approach, but with even greater variation. The correlation of tracer velocity at $\textrm{St}$=0 is slightly over-predicted at small scales and slightly under-predicted at large scales. Comparable dissimilarity is observed between the DNS and model at $\textrm{St}$=0.1. As the Stokes number increases, the linear filter of the model breaks down and the predictions decorrelate slower when compared to the simulated data responses.
\begin{figure}
  \centering
  \includegraphics[width=0.49 \textwidth]{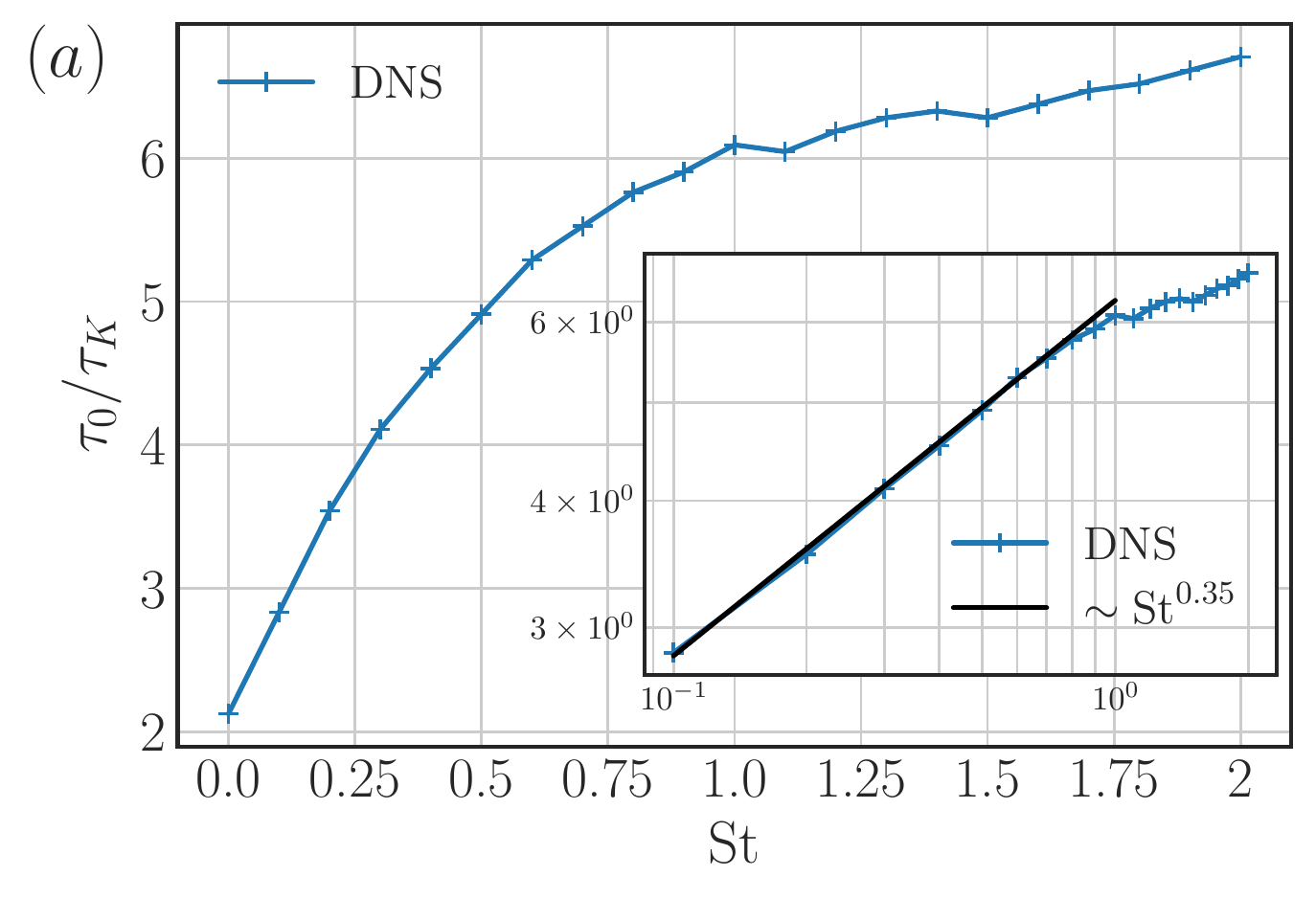}
  \includegraphics[width=0.49 \textwidth]{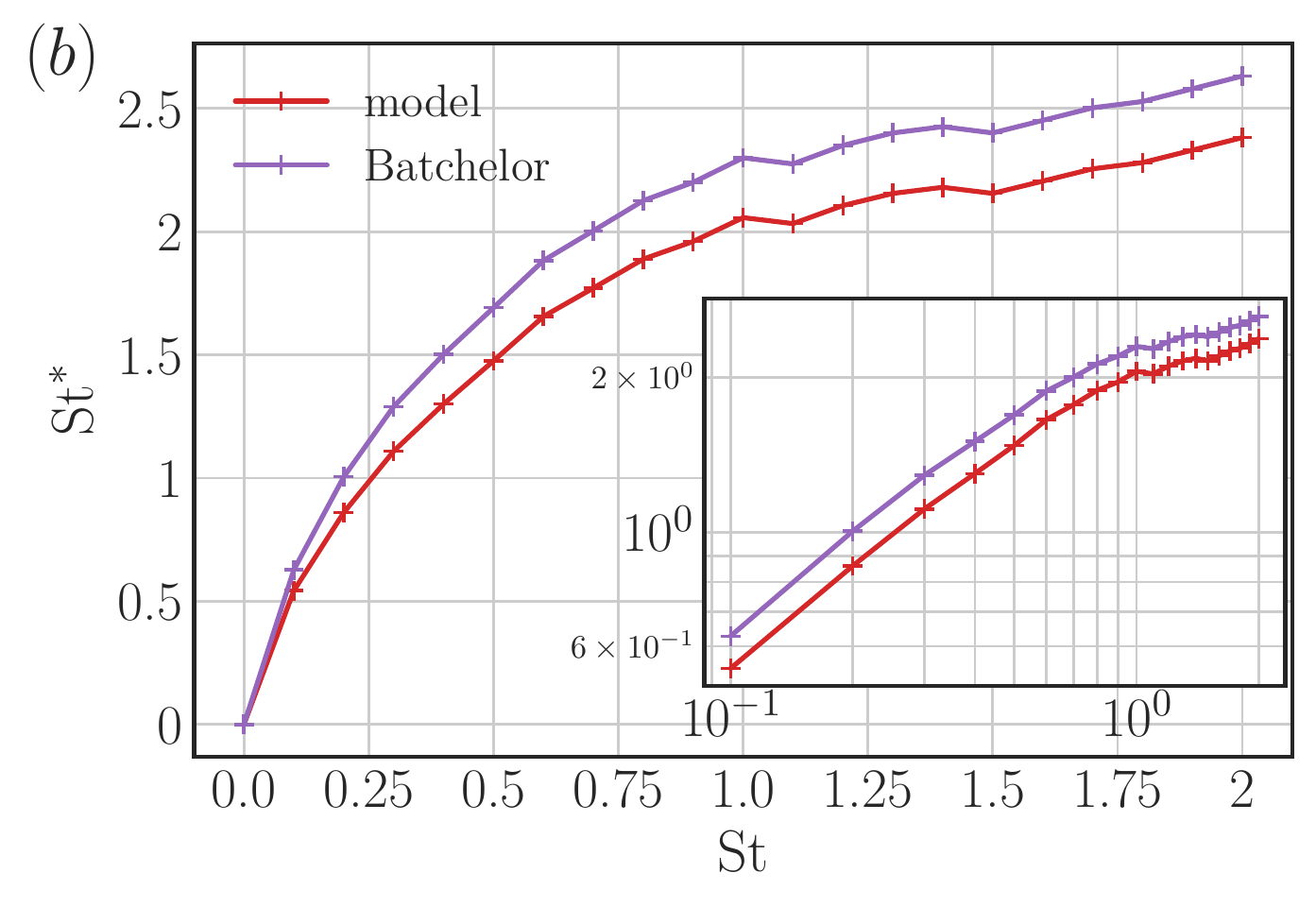}
  \caption{(a) Zero crossing $\tau_0$ of the acceleration correlation functions from DNS (as partially shown in Figure~\ref{fig:acc_corr_filtering}) as a function of $\textrm{St}$. The inset shows a double-logarithmic representation with the black line indicating $\tau_0 \sim \textrm{St}^{0.35}$. (b) Recalibrated particle response time $\tau_p^*$ expressed as $\textrm{St}^*=\tau_p^*/\tau_K$ based on the zero crossing $\tau_0$ (see main text for more details on the calibration process) as a function of $\textrm{St}$.}
  \label{fig:zero_cross}
\end{figure}

Due to the fact that the model calibration for finite $\textrm{St}$ is based on the zero crossing  of the acceleration correlation, it is worth studying the $\textrm{St}$-dependence of this quantity as well. Therefore, Figure~\ref{fig:zero_cross}(a) depicts the zero crossing $\tau_0$ from DNS as
a function of the Stokes number $\textrm{St}$. For the case of Lagrangian tracers $\textrm{St}=0$, the zero crossing is at around $2.2\tau_K$ and increases sharply for $\textrm{St}>0$. For $\textrm{St}\approx 1$, a new quantitative behavior sets in, and the zero crossing exhibits a slower increase. Furthermore, the zero crossing becomes rather noisy, therefore, it is not entirely clear whether the zero crossing would saturate at even higher $\textrm{St}$. The inset of Figure~\ref{fig:zero_cross}(a) shows a double-logarithmic representation of $\tau_0$. For $\textrm{St}<1$, the zero crossing appears to be a power law, whereas deviations from this power law appear at $\textrm{St}\approx 1$.
For comparison, the black line shows a power law $\sim \textrm{St}^{0.35}$. However, at this point, no clear phenomenological description could be provided that would allow for the explanation of such a power law of the zeros of acceleration for inertial particle motion.

\begin{figure}
  \centering
  \includegraphics[width=0.49 \textwidth]{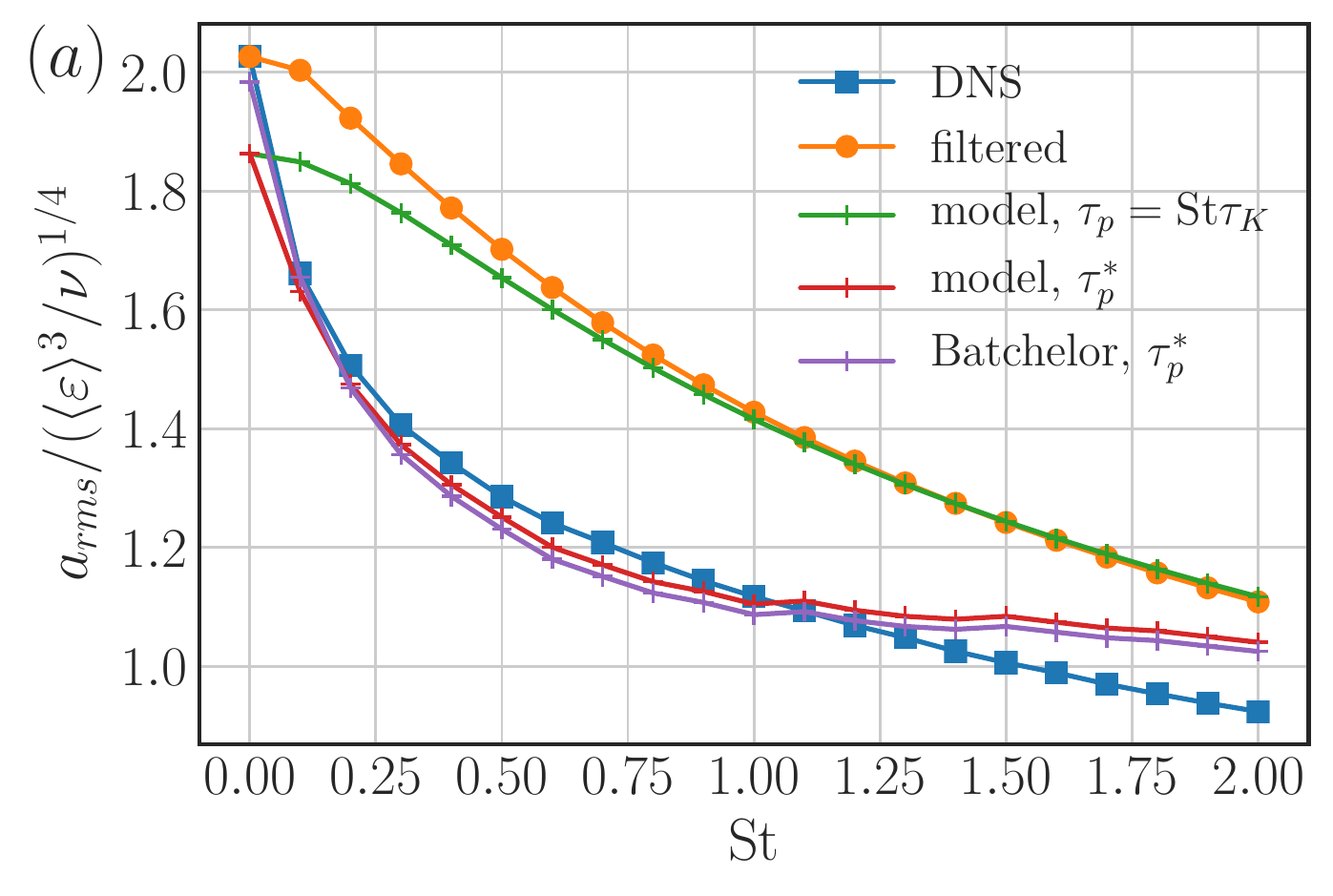}
  \includegraphics[width=0.49 \textwidth]{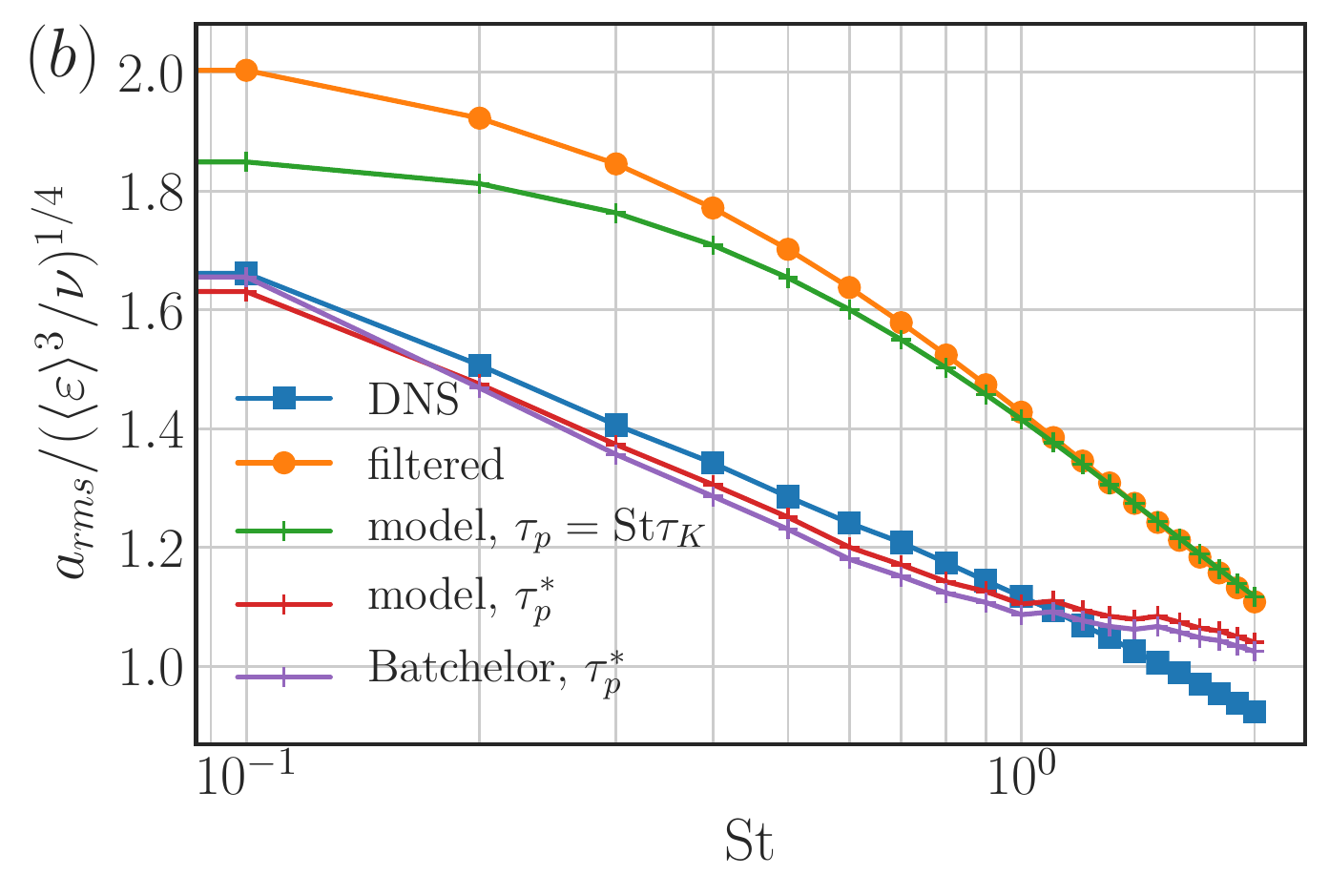}
  \caption{(a) Comparison of root mean square values of acceleration for DNS and the different models. (b) Semi-logarithmic representation of (a).}
  \label{fig:acc_rms}
\end{figure}

 Figure~\ref{fig:zero_cross}(b) depicts the calibrated effective Stokes number $\textrm{St}^*=\frac{\tau_p^*}{\tau_K}$ as a function of the DNS Stokes number $\textrm{St}$ for both the stochastic (red) and the Batchelor model (violet). Interestingly, the curves strongly resemble Figure~\ref{fig:zero_cross}(a)  which suggests a nearly linear relation between the zero crossing $\tau_0$ and the effective (calibrated) particle response time $\tau_p^*$.
The $a_{rms}$ values as a function of the Stokes number for all modeling techniques are included in Figure~\ref{fig:acc_rms} for direct comparison of the statistic. The filtered acceleration, found directly from Eq. (\ref{eq:LinkCapCa3}) (orange), and the stochastic model, based on $\tau_p$ (green), quickly deviate from the DNS $a_{rms}$ curve while the two models with the updated $\tau_p^*$ for the stochastic process (red) and Batchelor representation (purple) show improved agreement at all $\textrm{St}$ and good agreement between $0.2 \leq \textrm{St} \leq 1.1$.
Cencini $et\,al.$~ \cite{cencini2006dynamics} present a similar comparison and suggest that the deviation between the profiles at small Stokes numbers is due to preferential concentration captured which is not captured by the linear filter approximation. The inclusion of the effective particle response time $\tau_p^*$ in our model counteracts this discrepancy, providing accurate representations of acceleration statistics for the presented Stokes numbers.

%


\section{Conclusion}
\label{sec:conclusion}
We have presented a modeling technique for single inertial particle statistics based on a filtering approach for the Lagrangian fluid velocity. The introduction of an effective particle response time $\tau_p^*$ in the linear filter is motivated by the strong discrepancies in the dynamics of tracer-like and true inertial particles at low Stokes numbers. In particular, the effective particle response time accounts for the effects of preferential sampling of depleted vorticity regions by inertial particles, which manifests itself, for instance, in a strong decrease of the variance of the particle acceleration as suggested by Figure~\ref{fig:acc_rms}. Both the stochastic and the Batchelor model are capable of reproducing this characteristic feature of inertial particles. In contrast to the refinement of the linear filtering approximation proposed by Deutsch and Simonin~\cite{deutsch1992dispersion,deutsch1991large}, here, we do not operate on the level of fluid quantities as ``seen'' by inertial particles, but rather modify the ``response of particles to fluid structures''. This has the advantage that the calibration of our model equations is rather simple (i.e., it suffices to determine the empirically observed zero crossing of the acceleration correlation) in comparison to evaluating fluid quantities on the basis of individual inertial particle trajectories.
It will be a task for the future, to further investigate the peculiar notion of preferential sampling/preferential concentration by extending the modeling techniques and possibly combine them with generalized vortex models similar to those in Refs.~\cite{Friedrich2013,marcu:1995,Kohler2016,Ayyalasomayajula_2008} . Furthermore -  as we currently operate at the level of second order statistics -  much more could be understood by focusing for instance on the behavior of the flatness of velocity increment which highlights the implication of the existence of intermittent corrections. To include this behavior, we would first need to derive the action of the linear filtering at this higher level of statistics, and then to generalize both the stochastic process as well as the parametrization of Batchelor to include intermittent corrections, as it was done for Lagrangian velocity in Ref.~\cite{Viggiano_2020}.
Furthermore, the proposed model might be generalized to particle pairs which opens up the way for investigating particle collisions~\cite{Pumir2016} on the basis of simple stochastic models or the multifractal formalism.

\begin{acknowledgements}
J.F. acknowledges funding from the Humboldt Foundation within a Feodor-Lynen fellowship and also benefited from financial support of the Project IDEXLYON of the University of Lyon in the framework of the French program ``Programme Investissements d'Avenir'' (ANR-16-IDEX-0005). B.V. and R.B.C. are supported by U.S. National Science Foundation grant (NSF-GEO-1756259). R.B.C. is also thankful for the support provided through the Fulbright Scholar Program.
\end{acknowledgements}

\bibliography{inertial_process.bib}

\begin{thebibliography}{44}%
\makeatletter
\providecommand \@ifxundefined [1]{%
 \@ifx{#1\undefined}
}%
\providecommand \@ifnum [1]{%
 \ifnum #1\expandafter \@firstoftwo
 \else \expandafter \@secondoftwo
 \fi
}%
\providecommand \@ifx [1]{%
 \ifx #1\expandafter \@firstoftwo
 \else \expandafter \@secondoftwo
 \fi
}%
\providecommand \natexlab [1]{#1}%
\providecommand \enquote  [1]{``#1''}%
\providecommand \bibnamefont  [1]{#1}%
\providecommand \bibfnamefont [1]{#1}%
\providecommand \citenamefont [1]{#1}%
\providecommand \href@noop [0]{\@secondoftwo}%
\providecommand \href [0]{\begingroup \@sanitize@url \@href}%
\providecommand \@href[1]{\@@startlink{#1}\@@href}%
\providecommand \@@href[1]{\endgroup#1\@@endlink}%
\providecommand \@sanitize@url [0]{\catcode `\\12\catcode `\$12\catcode
  `\&12\catcode `\#12\catcode `\^12\catcode `\_12\catcode `\%12\relax}%
\providecommand \@@startlink[1]{}%
\providecommand \@@endlink[0]{}%
\providecommand \url  [0]{\begingroup\@sanitize@url \@url }%
\providecommand \@url [1]{\endgroup\@href {#1}{\urlprefix }}%
\providecommand \urlprefix  [0]{URL }%
\providecommand \Eprint [0]{\href }%
\providecommand \doibase [0]{https://doi.org/}%
\providecommand \selectlanguage [0]{\@gobble}%
\providecommand \bibinfo  [0]{\@secondoftwo}%
\providecommand \bibfield  [0]{\@secondoftwo}%
\providecommand \translation [1]{[#1]}%
\providecommand \BibitemOpen [0]{}%
\providecommand \bibitemStop [0]{}%
\providecommand \bibitemNoStop [0]{.\EOS\space}%
\providecommand \EOS [0]{\spacefactor3000\relax}%
\providecommand \BibitemShut  [1]{\csname bibitem#1\endcsname}%
\let\auto@bib@innerbib\@empty
\bibitem [{\citenamefont {Monin}\ and\ \citenamefont {Yaglom}(2007)}]{monin}%
  \BibitemOpen
  \bibfield  {author} {\bibinfo {author} {\bibfnamefont {A.~S.}\ \bibnamefont
  {Monin}}\ and\ \bibinfo {author} {\bibfnamefont {A.~M.}\ \bibnamefont
  {Yaglom}},\ }\href@noop {} {\emph {\bibinfo {title} {{Statistical Fluid
  Mechanics: Mechanics of Turbulence}}}}\ (\bibinfo  {publisher} {Courier Dover
  Publications},\ \bibinfo {year} {2007})\BibitemShut {NoStop}%
\bibitem [{\citenamefont {Frisch}(1995)}]{frisch:1995}%
  \BibitemOpen
  \bibfield  {author} {\bibinfo {author} {\bibfnamefont {U.}~\bibnamefont
  {Frisch}},\ }\href@noop {} {\emph {\bibinfo {title} {{Turbulence}}}}\
  (\bibinfo  {publisher} {Cambridge University Press},\ \bibinfo {year}
  {1995})\BibitemShut {NoStop}%
\bibitem [{\citenamefont {Pope}(2000)}]{Pope}%
  \BibitemOpen
  \bibfield  {author} {\bibinfo {author} {\bibfnamefont {S.~B.}\ \bibnamefont
  {Pope}},\ }\href@noop {} {\emph {\bibinfo {title} {{Turbulent Flows}}}}\
  (\bibinfo  {publisher} {Cambridge University Press},\ \bibinfo {year}
  {2000})\BibitemShut {NoStop}%
\bibitem [{\citenamefont {Toschi}\ and\ \citenamefont
  {Bodenschatz}(2009)}]{toschi2009lagrangian}%
  \BibitemOpen
  \bibfield  {author} {\bibinfo {author} {\bibfnamefont {F.}~\bibnamefont
  {Toschi}}\ and\ \bibinfo {author} {\bibfnamefont {E.}~\bibnamefont
  {Bodenschatz}},\ }\bibfield  {title} {\bibinfo {title} {Lagrangian properties
  of particles in turbulence},\ }\href@noop {} {\bibfield  {journal} {\bibinfo
  {journal} {Annu. Rev. Fluid Mech.}\ }\textbf {\bibinfo {volume} {41}},\
  \bibinfo {pages} {375} (\bibinfo {year} {2009})}\BibitemShut {NoStop}%
\bibitem [{\citenamefont {Machicoane}\ \emph {et~al.}(2019)\citenamefont
  {Machicoane}, \citenamefont {Huck}, \citenamefont {Clark}, \citenamefont
  {Aliseda}, \citenamefont {Volk},\ and\ \citenamefont
  {Bourgoin}}]{Machicoane2019}%
  \BibitemOpen
  \bibfield  {author} {\bibinfo {author} {\bibfnamefont {N.}~\bibnamefont
  {Machicoane}}, \bibinfo {author} {\bibfnamefont {P.~D.}\ \bibnamefont
  {Huck}}, \bibinfo {author} {\bibfnamefont {A.}~\bibnamefont {Clark}},
  \bibinfo {author} {\bibfnamefont {A.}~\bibnamefont {Aliseda}}, \bibinfo
  {author} {\bibfnamefont {R.}~\bibnamefont {Volk}},\ and\ \bibinfo {author}
  {\bibfnamefont {M.}~\bibnamefont {Bourgoin}},\ }\bibfield  {title} {\bibinfo
  {title} {Recent developments in particle tracking diagnostics for turbulence
  research},\ }in\ \href@noop {} {\emph {\bibinfo {booktitle} {Flowing
  Matter}}}\ (\bibinfo  {publisher} {Springer},\ \bibinfo {year} {2019})\ pp.\
  \bibinfo {pages} {177--209}\BibitemShut {NoStop}%
\bibitem [{\citenamefont {Qureshi}\ \emph {et~al.}(2008)\citenamefont
  {Qureshi}, \citenamefont {Arrieta}, \citenamefont {Baudet}, \citenamefont
  {Cartellier}, \citenamefont {Gagne},\ and\ \citenamefont
  {Bourgoin}}]{qureshi2008acceleration}%
  \BibitemOpen
  \bibfield  {author} {\bibinfo {author} {\bibfnamefont {N.~M.}\ \bibnamefont
  {Qureshi}}, \bibinfo {author} {\bibfnamefont {U.}~\bibnamefont {Arrieta}},
  \bibinfo {author} {\bibfnamefont {C.}~\bibnamefont {Baudet}}, \bibinfo
  {author} {\bibfnamefont {A.}~\bibnamefont {Cartellier}}, \bibinfo {author}
  {\bibfnamefont {Y.}~\bibnamefont {Gagne}},\ and\ \bibinfo {author}
  {\bibfnamefont {M.}~\bibnamefont {Bourgoin}},\ }\bibfield  {title} {\bibinfo
  {title} {{Acceleration statistics of inertial particles in turbulent flow}},\
  }\href@noop {} {\bibfield  {journal} {\bibinfo  {journal} {Eur. Phys. J. B}\
  }\textbf {\bibinfo {volume} {66}},\ \bibinfo {pages} {531} (\bibinfo {year}
  {2008})}\BibitemShut {NoStop}%
\bibitem [{\citenamefont {Ayyalasomayajula}\ \emph {et~al.}(2006)\citenamefont
  {Ayyalasomayajula}, \citenamefont {Gylfason}, \citenamefont {Collins},
  \citenamefont {Bodenschatz},\ and\ \citenamefont
  {Warhaft}}]{ayyalasomayajula2006lagrangian}%
  \BibitemOpen
  \bibfield  {author} {\bibinfo {author} {\bibfnamefont {S.}~\bibnamefont
  {Ayyalasomayajula}}, \bibinfo {author} {\bibfnamefont {A.}~\bibnamefont
  {Gylfason}}, \bibinfo {author} {\bibfnamefont {L.~R.}\ \bibnamefont
  {Collins}}, \bibinfo {author} {\bibfnamefont {E.}~\bibnamefont
  {Bodenschatz}},\ and\ \bibinfo {author} {\bibfnamefont {Z.}~\bibnamefont
  {Warhaft}},\ }\bibfield  {title} {\bibinfo {title} {{Lagrangian measurements
  of inertial particle accelerations in grid generated wind tunnel
  turbulence}},\ }\href@noop {} {\bibfield  {journal} {\bibinfo  {journal}
  {Phys. Rev. Lett.}\ }\textbf {\bibinfo {volume} {97}},\ \bibinfo {pages}
  {144507} (\bibinfo {year} {2006})}\BibitemShut {NoStop}%
\bibitem [{\citenamefont {Squires}\ and\ \citenamefont
  {Eaton}(1991)}]{Squires1991}%
  \BibitemOpen
  \bibfield  {author} {\bibinfo {author} {\bibfnamefont {K.~D.}\ \bibnamefont
  {Squires}}\ and\ \bibinfo {author} {\bibfnamefont {J.~K.}\ \bibnamefont
  {Eaton}},\ }\bibfield  {title} {\bibinfo {title} {Preferential concentration
  of particles by turbulence},\ }\href {https://doi.org/10.1063/1.858045}
  {\bibfield  {journal} {\bibinfo  {journal} {Phys. Fluids}\ }\textbf {\bibinfo
  {volume} {3}},\ \bibinfo {pages} {1169} (\bibinfo {year} {1991})}\BibitemShut
  {NoStop}%
\bibitem [{\citenamefont {Eaton}\ and\ \citenamefont
  {Fessler}(1994)}]{Eaton1994}%
  \BibitemOpen
  \bibfield  {author} {\bibinfo {author} {\bibfnamefont {J.~K.}\ \bibnamefont
  {Eaton}}\ and\ \bibinfo {author} {\bibfnamefont {J.~R.}\ \bibnamefont
  {Fessler}},\ }\bibfield  {title} {\bibinfo {title} {Preferential
  concentration of particles by turbulence},\ }\href@noop {} {\bibfield
  {journal} {\bibinfo  {journal} {Int. J. Multiph. Flow}\ }\textbf {\bibinfo
  {volume} {20}},\ \bibinfo {pages} {169} (\bibinfo {year} {1994})}\BibitemShut
  {NoStop}%
\bibitem [{\citenamefont {Monchaux}\ \emph {et~al.}(2012)\citenamefont
  {Monchaux}, \citenamefont {Bourgoin},\ and\ \citenamefont
  {Cartellier}}]{Monchaux2012}%
  \BibitemOpen
  \bibfield  {author} {\bibinfo {author} {\bibfnamefont {R.}~\bibnamefont
  {Monchaux}}, \bibinfo {author} {\bibfnamefont {M.}~\bibnamefont {Bourgoin}},\
  and\ \bibinfo {author} {\bibfnamefont {A.}~\bibnamefont {Cartellier}},\
  }\bibfield  {title} {\bibinfo {title} {{Analyzing preferential concentration
  and clustering of inertial particles in turbulence}},\ }\href@noop {}
  {\bibfield  {journal} {\bibinfo  {journal} {Int. J. Multiph. Flow}\ }\textbf
  {\bibinfo {volume} {40}},\ \bibinfo {pages} {1} (\bibinfo {year}
  {2012})}\BibitemShut {NoStop}%
\bibitem [{\citenamefont {Bourgoin}\ and\ \citenamefont
  {Xu}(2014)}]{Bourgoin2014}%
  \BibitemOpen
  \bibfield  {author} {\bibinfo {author} {\bibfnamefont {M.}~\bibnamefont
  {Bourgoin}}\ and\ \bibinfo {author} {\bibfnamefont {H.}~\bibnamefont {Xu}},\
  }\bibfield  {title} {\bibinfo {title} {Focus on dynamics of particles in
  turbulence},\ }\href {https://doi.org/10.1088/1367-2630/16/8/085010}
  {\bibfield  {journal} {\bibinfo  {journal} {New J. Phys.}\ }\textbf {\bibinfo
  {volume} {16}},\ \bibinfo {pages} {085010} (\bibinfo {year}
  {2014})}\BibitemShut {NoStop}%
\bibitem [{\citenamefont {Maxey}\ and\ \citenamefont
  {Riley}(1983)}]{Maxey1983}%
  \BibitemOpen
  \bibfield  {author} {\bibinfo {author} {\bibfnamefont {M.~R.}\ \bibnamefont
  {Maxey}}\ and\ \bibinfo {author} {\bibfnamefont {J.~J.}\ \bibnamefont
  {Riley}},\ }\bibfield  {title} {\bibinfo {title} {{Equation of motion for a
  small rigid sphere in a nonuniform flow}},\ }\href
  {https://doi.org/10.1063/1.864230} {\bibfield  {journal} {\bibinfo  {journal}
  {Phys. Fluids}\ }\textbf {\bibinfo {volume} {26}},\ \bibinfo {pages} {883}
  (\bibinfo {year} {1983})}\BibitemShut {NoStop}%
\bibitem [{\citenamefont {Chen}\ \emph {et~al.}(2006)\citenamefont {Chen},
  \citenamefont {Goto},\ and\ \citenamefont {Vassilicos}}]{Chen2006}%
  \BibitemOpen
  \bibfield  {author} {\bibinfo {author} {\bibfnamefont {L.}~\bibnamefont
  {Chen}}, \bibinfo {author} {\bibfnamefont {S.}~\bibnamefont {Goto}},\ and\
  \bibinfo {author} {\bibfnamefont {J.~C.}\ \bibnamefont {Vassilicos}},\
  }\bibfield  {title} {\bibinfo {title} {Turbulent clustering of stagnation
  points and inertial particles},\ }\href
  {https://doi.org/10.1017/S0022112006009177} {\bibfield  {journal} {\bibinfo
  {journal} {J. Fluid Mech.}\ }\textbf {\bibinfo {volume} {553}},\ \bibinfo
  {pages} {143} (\bibinfo {year} {2006})}\BibitemShut {NoStop}%
\bibitem [{\citenamefont {Gustavsson}\ and\ \citenamefont
  {Mehlig}(2016)}]{Gustavsson2016}%
  \BibitemOpen
  \bibfield  {author} {\bibinfo {author} {\bibfnamefont {K.}~\bibnamefont
  {Gustavsson}}\ and\ \bibinfo {author} {\bibfnamefont {B.}~\bibnamefont
  {Mehlig}},\ }\bibfield  {title} {\bibinfo {title} {Statistical models for
  spatial patterns of heavy particles in turbulence},\ }\href
  {https://doi.org/10.1080/00018732.2016.1164490} {\bibfield  {journal}
  {\bibinfo  {journal} {Advances in Physics}\ }\textbf {\bibinfo {volume}
  {65}},\ \bibinfo {pages} {1} (\bibinfo {year} {2016})}\BibitemShut {NoStop}%
\bibitem [{\citenamefont {Bec}\ \emph {et~al.}(2006)\citenamefont {Bec},
  \citenamefont {Biferale}, \citenamefont {Boffetta}, \citenamefont {Celani},
  \citenamefont {Cencini}, \citenamefont {Lanotte}, \citenamefont {Musacchio},\
  and\ \citenamefont {Toschi}}]{bec2006acceleration}%
  \BibitemOpen
  \bibfield  {author} {\bibinfo {author} {\bibfnamefont {J.}~\bibnamefont
  {Bec}}, \bibinfo {author} {\bibfnamefont {L.}~\bibnamefont {Biferale}},
  \bibinfo {author} {\bibfnamefont {G.}~\bibnamefont {Boffetta}}, \bibinfo
  {author} {\bibfnamefont {A.}~\bibnamefont {Celani}}, \bibinfo {author}
  {\bibfnamefont {M.}~\bibnamefont {Cencini}}, \bibinfo {author} {\bibfnamefont
  {A.}~\bibnamefont {Lanotte}}, \bibinfo {author} {\bibfnamefont
  {S.}~\bibnamefont {Musacchio}},\ and\ \bibinfo {author} {\bibfnamefont
  {F.}~\bibnamefont {Toschi}},\ }\bibfield  {title} {\bibinfo {title}
  {{Acceleration statistics of heavy particles in turbulence}},\ }\href@noop {}
  {\bibfield  {journal} {\bibinfo  {journal} {J. Fluid Mech.}\ }\textbf
  {\bibinfo {volume} {550}},\ \bibinfo {pages} {349} (\bibinfo {year}
  {2006})}\BibitemShut {NoStop}%
\bibitem [{\citenamefont {Bourgoin}\ \emph {et~al.}(2006)\citenamefont
  {Bourgoin}, \citenamefont {Ouellette}, \citenamefont {Xu}, \citenamefont
  {Berg},\ and\ \citenamefont {Bodenschatz}}]{bourgoin2006role}%
  \BibitemOpen
  \bibfield  {author} {\bibinfo {author} {\bibfnamefont {M.}~\bibnamefont
  {Bourgoin}}, \bibinfo {author} {\bibfnamefont {N.~T.}\ \bibnamefont
  {Ouellette}}, \bibinfo {author} {\bibfnamefont {H.}~\bibnamefont {Xu}},
  \bibinfo {author} {\bibfnamefont {J.}~\bibnamefont {Berg}},\ and\ \bibinfo
  {author} {\bibfnamefont {E.}~\bibnamefont {Bodenschatz}},\ }\bibfield
  {title} {\bibinfo {title} {The role of pair dispersion in turbulent flow},\
  }\href@noop {} {\bibfield  {journal} {\bibinfo  {journal} {Science}\ }\textbf
  {\bibinfo {volume} {311}},\ \bibinfo {pages} {835} (\bibinfo {year}
  {2006})}\BibitemShut {NoStop}%
\bibitem [{\citenamefont {Taylor}(1922)}]{taylor:1922}%
  \BibitemOpen
  \bibfield  {author} {\bibinfo {author} {\bibfnamefont {G.~I.}\ \bibnamefont
  {Taylor}},\ }\bibfield  {title} {\bibinfo {title} {{Diffusion by Continuous
  Movements}},\ }\href {https://doi.org/10.1112/plms/s2-20.1.196} {\bibfield
  {journal} {\bibinfo  {journal} {Proc. London Math. Soc.}\ }\textbf {\bibinfo
  {volume} {s2-20}},\ \bibinfo {pages} {196} (\bibinfo {year}
  {1922})}\BibitemShut {NoStop}%
\bibitem [{\citenamefont {Obukhov}(1959)}]{OBUKHOV1959113}%
  \BibitemOpen
  \bibfield  {author} {\bibinfo {author} {\bibfnamefont {A.}~\bibnamefont
  {Obukhov}},\ }\bibfield  {title} {\bibinfo {title} {Description of turbulence
  in terms of lagrangian variables}\ }(\bibinfo  {publisher} {Elsevier},\
  \bibinfo {year} {1959})\ pp.\ \bibinfo {pages} {113--116}\BibitemShut
  {NoStop}%
\bibitem [{\citenamefont {Sawford}(1991)}]{sawford1991reynolds}%
  \BibitemOpen
  \bibfield  {author} {\bibinfo {author} {\bibfnamefont {B.~L.}\ \bibnamefont
  {Sawford}},\ }\bibfield  {title} {\bibinfo {title} {{Reynolds number effects
  in Lagrangian stochastic models of turbulent dispersion}},\ }\href@noop {}
  {\bibfield  {journal} {\bibinfo  {journal} {Phys. Fluids}\ }\textbf {\bibinfo
  {volume} {3}},\ \bibinfo {pages} {1577} (\bibinfo {year} {1991})}\BibitemShut
  {NoStop}%
\bibitem [{\citenamefont {Pinton}\ and\ \citenamefont {Sawford}()}]{Pinton}%
  \BibitemOpen
  \bibfield  {author} {\bibinfo {author} {\bibfnamefont {J.-F.}\ \bibnamefont
  {Pinton}}\ and\ \bibinfo {author} {\bibfnamefont {B.~L.}\ \bibnamefont
  {Sawford}},\ }\bibfield  {title} {\bibinfo {title} {A lagrangian view of
  turbulent dispersion and mixing},\ }in\ \href
  {https://doi.org/10.1017/cbo9781139032810.005} {\emph {\bibinfo {booktitle}
  {Ten Chapters in Turbulence}}},\ \bibinfo {editor} {edited by\ \bibinfo
  {editor} {\bibfnamefont {P.~A.}\ \bibnamefont {Davidson}}, \bibinfo {editor}
  {\bibfnamefont {Y.}~\bibnamefont {Kaneda}},\ and\ \bibinfo {editor}
  {\bibfnamefont {K.~R.}\ \bibnamefont {Sreenivasan}}}\ (\bibinfo  {publisher}
  {Cambridge University Press})\ pp.\ \bibinfo {pages} {132--175}\BibitemShut
  {NoStop}%
\bibitem [{\citenamefont {Viggiano}\ \emph {et~al.}(2020)\citenamefont
  {Viggiano}, \citenamefont {Friedrich}, \citenamefont {Volk}, \citenamefont
  {Bourgoin}, \citenamefont {Cal},\ and\ \citenamefont
  {Chevillard}}]{Viggiano_2020}%
  \BibitemOpen
  \bibfield  {author} {\bibinfo {author} {\bibfnamefont {B.}~\bibnamefont
  {Viggiano}}, \bibinfo {author} {\bibfnamefont {J.}~\bibnamefont {Friedrich}},
  \bibinfo {author} {\bibfnamefont {R.}~\bibnamefont {Volk}}, \bibinfo {author}
  {\bibfnamefont {M.}~\bibnamefont {Bourgoin}}, \bibinfo {author}
  {\bibfnamefont {R.~B.}\ \bibnamefont {Cal}},\ and\ \bibinfo {author}
  {\bibfnamefont {L.}~\bibnamefont {Chevillard}},\ }\bibfield  {title}
  {\bibinfo {title} {{Modelling Lagrangian velocity and acceleration in
  turbulent flows as infinitely differentiable stochastic processes}},\ }\href
  {https://doi.org/10.1017/jfm.2020.495} {\bibfield  {journal} {\bibinfo
  {journal} {J. Fluid Mech.}\ }\textbf {\bibinfo {volume} {900}},\ \bibinfo
  {pages} {A27} (\bibinfo {year} {2020})}\BibitemShut {NoStop}%
\bibitem [{\citenamefont {Chevillard}\ \emph {et~al.}(2003)\citenamefont
  {Chevillard}, \citenamefont {Roux}, \citenamefont {L{\'e}v{\^e}que},
  \citenamefont {Mordant}, \citenamefont {Pinton},\ and\ \citenamefont
  {Arn{\'e}odo}}]{chevillard2003lagrangian}%
  \BibitemOpen
  \bibfield  {author} {\bibinfo {author} {\bibfnamefont {L.}~\bibnamefont
  {Chevillard}}, \bibinfo {author} {\bibfnamefont {S.}~\bibnamefont {Roux}},
  \bibinfo {author} {\bibfnamefont {E.}~\bibnamefont {L{\'e}v{\^e}que}},
  \bibinfo {author} {\bibfnamefont {N.}~\bibnamefont {Mordant}}, \bibinfo
  {author} {\bibfnamefont {J.-F.}\ \bibnamefont {Pinton}},\ and\ \bibinfo
  {author} {\bibfnamefont {A.}~\bibnamefont {Arn{\'e}odo}},\ }\bibfield
  {title} {\bibinfo {title} {Lagrangian velocity statistics in turbulent flows:
  Effects of dissipation},\ }\href@noop {} {\bibfield  {journal} {\bibinfo
  {journal} {Phys. Rev. Lett.}\ }\textbf {\bibinfo {volume} {91}},\ \bibinfo
  {pages} {214502} (\bibinfo {year} {2003})}\BibitemShut {NoStop}%
\bibitem [{\citenamefont {Biferale}\ \emph {et~al.}(2004)\citenamefont
  {Biferale}, \citenamefont {Boffetta}, \citenamefont {Celani}, \citenamefont
  {Devenish}, \citenamefont {Lanotte},\ and\ \citenamefont
  {Toschi}}]{biferale2004multifractal}%
  \BibitemOpen
  \bibfield  {author} {\bibinfo {author} {\bibfnamefont {L.}~\bibnamefont
  {Biferale}}, \bibinfo {author} {\bibfnamefont {G.}~\bibnamefont {Boffetta}},
  \bibinfo {author} {\bibfnamefont {A.}~\bibnamefont {Celani}}, \bibinfo
  {author} {\bibfnamefont {B.}~\bibnamefont {Devenish}}, \bibinfo {author}
  {\bibfnamefont {A.}~\bibnamefont {Lanotte}},\ and\ \bibinfo {author}
  {\bibfnamefont {F.}~\bibnamefont {Toschi}},\ }\bibfield  {title} {\bibinfo
  {title} {Multifractal statistics of lagrangian velocity and acceleration in
  turbulence},\ }\href@noop {} {\bibfield  {journal} {\bibinfo  {journal}
  {Phys. Rev. Lett.}\ }\textbf {\bibinfo {volume} {93}},\ \bibinfo {pages}
  {064502} (\bibinfo {year} {2004})}\BibitemShut {NoStop}%
\bibitem [{\citenamefont {Arn{\'{e}}odo~et al.}(2008)}]{arneodo2008universal}%
  \BibitemOpen
  \bibfield  {author} {\bibinfo {author} {\bibfnamefont {A.}~\bibnamefont
  {Arn{\'{e}}odo~et al.}},\ }\bibfield  {title} {\bibinfo {title} {Universal
  intermittent properties of particle trajectories in highly turbulent flows},\
  }\href@noop {} {\bibfield  {journal} {\bibinfo  {journal} {Phys. Rev. Lett.}\
  }\textbf {\bibinfo {volume} {100}},\ \bibinfo {pages} {254504} (\bibinfo
  {year} {2008})}\BibitemShut {NoStop}%
\bibitem [{\citenamefont {Chevillard}\ \emph {et~al.}(2012)\citenamefont
  {Chevillard}, \citenamefont {Castaing}, \citenamefont {Arneodo},
  \citenamefont {L{\'e}v{\^e}que}, \citenamefont {Pinton},\ and\ \citenamefont
  {Roux}}]{chevillard2012phenomenological}%
  \BibitemOpen
  \bibfield  {author} {\bibinfo {author} {\bibfnamefont {L.}~\bibnamefont
  {Chevillard}}, \bibinfo {author} {\bibfnamefont {B.}~\bibnamefont
  {Castaing}}, \bibinfo {author} {\bibfnamefont {A.}~\bibnamefont {Arneodo}},
  \bibinfo {author} {\bibfnamefont {E.}~\bibnamefont {L{\'e}v{\^e}que}},
  \bibinfo {author} {\bibfnamefont {J.-F.}\ \bibnamefont {Pinton}},\ and\
  \bibinfo {author} {\bibfnamefont {S.~G.}\ \bibnamefont {Roux}},\ }\bibfield
  {title} {\bibinfo {title} {A phenomenological theory of {E}ulerian and
  {L}agrangian velocity fluctuations in turbulent flows},\ }\href@noop {}
  {\bibfield  {journal} {\bibinfo  {journal} {C R Phys}\ }\textbf {\bibinfo
  {volume} {13}},\ \bibinfo {pages} {899} (\bibinfo {year} {2012})}\BibitemShut
  {NoStop}%
\bibitem [{\citenamefont {Friedrich}(2003)}]{friedrich:2003}%
  \BibitemOpen
  \bibfield  {author} {\bibinfo {author} {\bibfnamefont {R.}~\bibnamefont
  {Friedrich}},\ }\bibfield  {title} {\bibinfo {title} {Statistics of
  lagrangian velocities in turbulent flows},\ }\href
  {https://doi.org/10.1103/PhysRevLett.90.084501} {\bibfield  {journal}
  {\bibinfo  {journal} {Phys. Rev. Lett.}\ }\textbf {\bibinfo {volume} {90}},\
  \bibinfo {pages} {84501} (\bibinfo {year} {2003})}\BibitemShut {NoStop}%
\bibitem [{\citenamefont {Gatignol}(1983)}]{Gatignol}%
  \BibitemOpen
  \bibfield  {author} {\bibinfo {author} {\bibfnamefont {R.}~\bibnamefont
  {Gatignol}},\ }\bibfield  {title} {\bibinfo {title} {The faxen formulas for a
  rigid particle in an unsteady non-uniform {S}tokes flow},\ }\href@noop {}
  {\bibfield  {journal} {\bibinfo  {journal} {J. Mec. Theor. Appl,}\ }\textbf
  {\bibinfo {volume} {1}},\ \bibinfo {pages} {143} (\bibinfo {year}
  {1983})}\BibitemShut {NoStop}%
\bibitem [{\citenamefont {Pumir}\ and\ \citenamefont
  {Wilkinson}(2016{\natexlab{a}})}]{Pumir2016}%
  \BibitemOpen
  \bibfield  {author} {\bibinfo {author} {\bibfnamefont {A.}~\bibnamefont
  {Pumir}}\ and\ \bibinfo {author} {\bibfnamefont {M.}~\bibnamefont
  {Wilkinson}},\ }\bibfield  {title} {\bibinfo {title} {{Collisional
  Aggregation Due to Turbulence}},\ }\href
  {https://doi.org/10.1146/annurev-conmatphys-031115-011538} {\bibfield
  {journal} {\bibinfo  {journal} {Annu. Rev. Condens. Matter Phys.}\ }\textbf
  {\bibinfo {volume} {7}},\ \bibinfo {pages} {141} (\bibinfo {year}
  {2016}{\natexlab{a}})}\BibitemShut {NoStop}%
\bibitem [{\citenamefont {Tchen}(2013)}]{chan2013mean}%
  \BibitemOpen
  \bibfield  {author} {\bibinfo {author} {\bibfnamefont {C.-M.}\ \bibnamefont
  {Tchen}},\ }\href@noop {} {\emph {\bibinfo {title} {Mean value and
  correlation problems connected with the motion of small particles suspended
  in a turbulent fluid}}}\ (\bibinfo  {publisher} {Springer},\ \bibinfo {year}
  {2013})\BibitemShut {NoStop}%
\bibitem [{\citenamefont {Hinze}(1959)}]{hinze1959turbulence}%
  \BibitemOpen
  \bibfield  {author} {\bibinfo {author} {\bibfnamefont {J.}~\bibnamefont
  {Hinze}},\ }\bibfield  {title} {\bibinfo {title} {Turbulence mcgraw-hill},\
  }\href@noop {} {\bibfield  {journal} {\bibinfo  {journal} {New York}\
  }\textbf {\bibinfo {volume} {421}},\ \bibinfo {pages} {3} (\bibinfo {year}
  {1959})}\BibitemShut {NoStop}%
\bibitem [{\citenamefont {Deutsch}(1992)}]{deutsch1992dispersion}%
  \BibitemOpen
  \bibfield  {author} {\bibinfo {author} {\bibfnamefont {E.}~\bibnamefont
  {Deutsch}},\ }\emph {\bibinfo {title} {Dispersion de particules dans une
  turbulence homog{\`e}ne isotrope stationnaire calcul{\'e}e par simulation
  num{\'e}rique directe des grandes {\'e}chelles}},\ \href@noop {} {Ph.D.
  thesis},\ \bibinfo  {school} {Ecully, Ecole centrale de Lyon} (\bibinfo
  {year} {1992})\BibitemShut {NoStop}%
\bibitem [{\citenamefont {Deutsch}\ and\ \citenamefont
  {Simonin}(1991)}]{deutsch1991large}%
  \BibitemOpen
  \bibfield  {author} {\bibinfo {author} {\bibfnamefont {E.}~\bibnamefont
  {Deutsch}}\ and\ \bibinfo {author} {\bibfnamefont {O.}~\bibnamefont
  {Simonin}},\ }\bibfield  {title} {\bibinfo {title} {Large eddy simulation
  applied to the modelling of particulate transport coefficients in turbulent
  two-phase flows},\ }in\ \href@noop {} {\emph {\bibinfo {booktitle} {8th
  Symposium on Turbulent Shear Flows, Volume 1}}},\ Vol.~\bibinfo {volume} {1}\
  (\bibinfo {year} {1991})\ pp.\ \bibinfo {pages}
  {10\_1\_1--10\_1\_6}\BibitemShut {NoStop}%
\bibitem [{\citenamefont {F{\'e}vrier}(2000)}]{fevrier2000etude}%
  \BibitemOpen
  \bibfield  {author} {\bibinfo {author} {\bibfnamefont {P.}~\bibnamefont
  {F{\'e}vrier}},\ }\emph {\bibinfo {title} {Etude num{\'e}rique des effets de
  concentration pr{\'e}f{\'e}rentielle et de corr{\'e}lation spatiale entre
  vitesses de particules solides en turbulence homogene isotrope
  stationnaire}},\ \href@noop {} {Ph.D. thesis},\ \bibinfo  {school} {Toulouse,
  INPT} (\bibinfo {year} {2000})\BibitemShut {NoStop}%
\bibitem [{\citenamefont {Berk}\ and\ \citenamefont
  {Coletti}(2021)}]{berk2021dynamics}%
  \BibitemOpen
  \bibfield  {author} {\bibinfo {author} {\bibfnamefont {T.}~\bibnamefont
  {Berk}}\ and\ \bibinfo {author} {\bibfnamefont {F.}~\bibnamefont {Coletti}},\
  }\bibfield  {title} {\bibinfo {title} {Dynamics of small heavy particles in
  homogeneous turbulence: a lagrangian experimental study},\ }\href@noop {}
  {\bibfield  {journal} {\bibinfo  {journal} {J. Fluid Mech.}\ }\textbf
  {\bibinfo {volume} {917}} (\bibinfo {year} {2021})}\BibitemShut {NoStop}%
\bibitem [{\citenamefont {Cencini}\ \emph {et~al.}(2006)\citenamefont
  {Cencini}, \citenamefont {Bec}, \citenamefont {Biferale}, \citenamefont
  {Boffetta}, \citenamefont {Celani}, \citenamefont {Lanotte}, \citenamefont
  {Musacchio},\ and\ \citenamefont {Toschi}}]{cencini2006dynamics}%
  \BibitemOpen
  \bibfield  {author} {\bibinfo {author} {\bibfnamefont {M.}~\bibnamefont
  {Cencini}}, \bibinfo {author} {\bibfnamefont {J.}~\bibnamefont {Bec}},
  \bibinfo {author} {\bibfnamefont {L.}~\bibnamefont {Biferale}}, \bibinfo
  {author} {\bibfnamefont {G.}~\bibnamefont {Boffetta}}, \bibinfo {author}
  {\bibfnamefont {A.}~\bibnamefont {Celani}}, \bibinfo {author} {\bibfnamefont
  {A.~S.}\ \bibnamefont {Lanotte}}, \bibinfo {author} {\bibfnamefont
  {S.}~\bibnamefont {Musacchio}},\ and\ \bibinfo {author} {\bibfnamefont
  {F.}~\bibnamefont {Toschi}},\ }\bibfield  {title} {\bibinfo {title}
  {{Dynamics and statistics of heavy particles in turbulent flows}},\
  }\href@noop {} {\bibfield  {journal} {\bibinfo  {journal} {J. Turbul.}\ ,\
  \bibinfo {pages} {N36}} (\bibinfo {year} {2006})}\BibitemShut {NoStop}%
\bibitem [{\citenamefont {Yu}\ \emph {et~al.}(2012)\citenamefont {Yu},
  \citenamefont {Kanov}, \citenamefont {Perlman}, \citenamefont {Graham},
  \citenamefont {Frederix}, \citenamefont {Burns}, \citenamefont {Szalay},
  \citenamefont {Eyink},\ and\ \citenamefont {Meneveau}}]{yu2012studying}%
  \BibitemOpen
  \bibfield  {author} {\bibinfo {author} {\bibfnamefont {H.}~\bibnamefont
  {Yu}}, \bibinfo {author} {\bibfnamefont {K.}~\bibnamefont {Kanov}}, \bibinfo
  {author} {\bibfnamefont {E.}~\bibnamefont {Perlman}}, \bibinfo {author}
  {\bibfnamefont {J.}~\bibnamefont {Graham}}, \bibinfo {author} {\bibfnamefont
  {E.}~\bibnamefont {Frederix}}, \bibinfo {author} {\bibfnamefont
  {R.}~\bibnamefont {Burns}}, \bibinfo {author} {\bibfnamefont
  {A.}~\bibnamefont {Szalay}}, \bibinfo {author} {\bibfnamefont
  {G.}~\bibnamefont {Eyink}},\ and\ \bibinfo {author} {\bibfnamefont
  {C.}~\bibnamefont {Meneveau}},\ }\bibfield  {title} {\bibinfo {title}
  {{Studying Lagrangian dynamics of turbulence using on-demand fluid particle
  tracking in a public turbulence database}},\ }\href@noop {} {\bibfield
  {journal} {\bibinfo  {journal} {J. Turbul.}\ ,\ \bibinfo {pages} {N12}}
  (\bibinfo {year} {2012})}\BibitemShut {NoStop}%
\bibitem [{\citenamefont {Pumir}\ and\ \citenamefont
  {Wilkinson}(2016{\natexlab{b}})}]{Pumir_2016}%
  \BibitemOpen
  \bibfield  {author} {\bibinfo {author} {\bibfnamefont {A.}~\bibnamefont
  {Pumir}}\ and\ \bibinfo {author} {\bibfnamefont {M.}~\bibnamefont
  {Wilkinson}},\ }\bibfield  {title} {\bibinfo {title} {{Collisional
  Aggregation Due to Turbulence}},\ }\href
  {https://doi.org/10.1146/annurev-conmatphys-031115-011538} {\bibfield
  {journal} {\bibinfo  {journal} {Annu. Rev. Condens. Matter Phys.}\ }\textbf
  {\bibinfo {volume} {7}},\ \bibinfo {pages} {141} (\bibinfo {year}
  {2016}{\natexlab{b}})}\BibitemShut {NoStop}%
\bibitem [{\citenamefont {Batchelor}(1951)}]{Batchelor1951}%
  \BibitemOpen
  \bibfield  {author} {\bibinfo {author} {\bibfnamefont {G.}~\bibnamefont
  {Batchelor}},\ }\bibfield  {title} {\bibinfo {title} {Pressure fluctuations
  in isotropic turbulence},\ }in\ \href@noop {} {\emph {\bibinfo {booktitle}
  {Math. Proc. Cambridge Philos. Soc.}}},\ Vol.~\bibinfo {volume} {47}\
  (\bibinfo {organization} {Cambridge University Press},\ \bibinfo {year}
  {1951})\ pp.\ \bibinfo {pages} {359--374}\BibitemShut {NoStop}%
\bibitem [{\citenamefont {Meneveau}(1996)}]{meneveau1996transition}%
  \BibitemOpen
  \bibfield  {author} {\bibinfo {author} {\bibfnamefont {C.}~\bibnamefont
  {Meneveau}},\ }\bibfield  {title} {\bibinfo {title} {Transition between
  viscous and inertial-range scaling of turbulence structure functions},\
  }\href@noop {} {\bibfield  {journal} {\bibinfo  {journal} {Phys. Rev. E}\
  }\textbf {\bibinfo {volume} {54}},\ \bibinfo {pages} {3657} (\bibinfo {year}
  {1996})}\BibitemShut {NoStop}%
\bibitem [{\citenamefont {Benzi}\ \emph {et~al.}(2010)\citenamefont {Benzi},
  \citenamefont {Biferale}, \citenamefont {Fisher}, \citenamefont {Lamb},\ and\
  \citenamefont {Toschi}}]{Benzi2010}%
  \BibitemOpen
  \bibfield  {author} {\bibinfo {author} {\bibfnamefont {R.}~\bibnamefont
  {Benzi}}, \bibinfo {author} {\bibfnamefont {L.}~\bibnamefont {Biferale}},
  \bibinfo {author} {\bibfnamefont {R.}~\bibnamefont {Fisher}}, \bibinfo
  {author} {\bibfnamefont {D.~Q.}\ \bibnamefont {Lamb}},\ and\ \bibinfo
  {author} {\bibfnamefont {F.}~\bibnamefont {Toschi}},\ }\bibfield  {title}
  {\bibinfo {title} {{Inertial range Eulerian and Lagrangian statistics from
  numerical simulations of isotropic turbulence}},\ }\href@noop {} {\bibfield
  {journal} {\bibinfo  {journal} {J. Fluid Mech.}\ }\textbf {\bibinfo {volume}
  {653}},\ \bibinfo {pages} {221 } (\bibinfo {year} {2010})}\BibitemShut
  {NoStop}%
\bibitem [{\citenamefont {Friedrich}\ and\ \citenamefont
  {Friedrich}(2013)}]{Friedrich2013}%
  \BibitemOpen
  \bibfield  {author} {\bibinfo {author} {\bibfnamefont {J.}~\bibnamefont
  {Friedrich}}\ and\ \bibinfo {author} {\bibfnamefont {R.}~\bibnamefont
  {Friedrich}},\ }\bibfield  {title} {\bibinfo {title} {{Generalized vortex
  model for the inverse cascade of two-dimensional turbulence}},\ }\href@noop
  {} {\bibfield  {journal} {\bibinfo  {journal} {Phys. Rev. E}\ }\textbf
  {\bibinfo {volume} {88}},\ \bibinfo {pages} {1} (\bibinfo {year}
  {2013})}\BibitemShut {NoStop}%
\bibitem [{\citenamefont {Marcu}\ \emph {et~al.}(1995)\citenamefont {Marcu},
  \citenamefont {Meiburg},\ and\ \citenamefont {Newton}}]{marcu:1995}%
  \BibitemOpen
  \bibfield  {author} {\bibinfo {author} {\bibfnamefont {B.}~\bibnamefont
  {Marcu}}, \bibinfo {author} {\bibfnamefont {E.}~\bibnamefont {Meiburg}},\
  and\ \bibinfo {author} {\bibfnamefont {P.~K.}\ \bibnamefont {Newton}},\
  }\bibfield  {title} {\bibinfo {title} {{Dynamics of heavy particles in a
  Burgers vortex}},\ }\href@noop {} {\bibfield  {journal} {\bibinfo  {journal}
  {Phys. Fluids}\ }\textbf {\bibinfo {volume} {7}},\ \bibinfo {pages} {400}
  (\bibinfo {year} {1995})}\BibitemShut {NoStop}%
\bibitem [{\citenamefont {K{\"{o}}hler}\ \emph {et~al.}(2016)\citenamefont
  {K{\"{o}}hler}, \citenamefont {Friedrich}, \citenamefont {Ostendorf},\ and\
  \citenamefont {Gurevich}}]{Kohler2016}%
  \BibitemOpen
  \bibfield  {author} {\bibinfo {author} {\bibfnamefont {J.}~\bibnamefont
  {K{\"{o}}hler}}, \bibinfo {author} {\bibfnamefont {J.}~\bibnamefont
  {Friedrich}}, \bibinfo {author} {\bibfnamefont {A.}~\bibnamefont
  {Ostendorf}},\ and\ \bibinfo {author} {\bibfnamefont {E.~L.}\ \bibnamefont
  {Gurevich}},\ }\bibfield  {title} {\bibinfo {title} {{Characterization of
  azimuthal and radial velocity fields induced by rotors in flows with a low
  Reynolds number}},\ }\href
  {http://link.aps.org/doi/10.1103/PhysRevE.93.023108} {\bibfield  {journal}
  {\bibinfo  {journal} {Phys. Rev. E}\ }\textbf {\bibinfo {volume} {93}},\
  \bibinfo {pages} {23108} (\bibinfo {year} {2016})}\BibitemShut {NoStop}%
\bibitem [{\citenamefont {Ayyalasomayajula}\ \emph {et~al.}(2008)\citenamefont
  {Ayyalasomayajula}, \citenamefont {Warhaft},\ and\ \citenamefont
  {Collins}}]{Ayyalasomayajula_2008}%
  \BibitemOpen
  \bibfield  {author} {\bibinfo {author} {\bibfnamefont {S.}~\bibnamefont
  {Ayyalasomayajula}}, \bibinfo {author} {\bibfnamefont {Z.}~\bibnamefont
  {Warhaft}},\ and\ \bibinfo {author} {\bibfnamefont {L.~R.}\ \bibnamefont
  {Collins}},\ }\bibfield  {title} {\bibinfo {title} {{Modeling inertial
  particle acceleration statistics in isotropic turbulence}},\ }\href
  {https://doi.org/10.1063/1.2976174} {\bibfield  {journal} {\bibinfo
  {journal} {Phys. Fluids}\ }\textbf {\bibinfo {volume} {20}},\ \bibinfo
  {pages} {95104} (\bibinfo {year} {2008})}\BibitemShut {NoStop}%
\end{thebibliography}%
\end{document}